\documentclass[aps,pra,twocolumn,final,showpacs,superscriptaddress]{revtex4}

\usepackage{amsmath}
\usepackage{amsfonts}
\usepackage{amssymb}
\usepackage{array}
\usepackage{float}
\usepackage[final]{epsfig}

\newcommand{\rmd}{\mathrm{d}}
\newcommand{\rme}{\mathrm{e}}
\newcommand{\rmi}{\mathrm{i}}
\newcommand{\Ai}{\operatorname{Ai}}
\newcommand{\Bi}{\operatorname{Bi}}
\newcommand{\Ci}{\operatorname{Ci}}
\newcommand{\imag}{\operatorname{Im}}

\begin{document}

\title{Electron dynamics in parallel electric and magnetic fields}
\author{Christian~Bracher}
\email{cbracher@brynmawr.edu}
\affiliation{Physics Department, Bryn Mawr College, Bryn Mawr, PA 19010, USA}
\author{Tobias Kramer}
\affiliation{Harvard University Department of Physics,
Cambridge, MA 02138, USA}
\email{tobias.kramer@mytum.de}
\author{John B.\ Delos}
\affiliation{Physics Department, The College of William and Mary, Williamsburg, VA 23187--8795, USA}
\email{jbdelo@wm.edu}
\date{\today}

\begin{abstract} 
We examine the spatial distribution of electrons generated by a fixed energy point source in uniform, parallel electric and magnetic fields.  This problem is simple enough to permit analytic quantum and semiclassical solution, and it harbors a rich set of features which find their interpretation in the unusual and interesting properties of the classical motion of the electrons: For instance, the number of interfering trajectories can be adjusted in this system, and the turning surfaces of classical motion contain a complex array of singularities.  We perform a comprehensive analysis of both the semiclassical approximation and the quantum solution, and we make predictions that should serve as a guide for future photodetachment experiments.
\end{abstract}

%

\maketitle

\section{Introduction}
\label{sec:Intro}

The photoabsorption cross-section of atoms or negative ions in presence of external electric \cite{Bryant1987a,Gibson1993a,Gibson2001a} or magnetic \cite{Blumberg1978a,Blumberg1979a} fields shows oscillations as a function of the photon energy.  For example, in a recent experiment, Yukich et al.\ \cite{Yukich2003a} measured the photodetachment rates of negative ions in a magnetic field and showed that the addition of a parallel electric field causes variations in the cross-section.  Closed Orbit Theory interprets the oscillations as interferences between waves that go out from, and return to, the atomic source \cite{Du1988a,Du1989a,Peters1994a,Peters1997a,Peters1997b}.  Such semiclassical models, as well as quantum calculations \cite{Blumberg1979a,Slonim1976a,Fabrikant1981a,Kondratovich1990a,Fabrikant1991a}, are in good agreement with experimental results. 

Many years ago, Demkov, Kondratovich and Ostrovskii \cite{Demkov1982a} suggested that interference structures could also be observed in a spatially resolved measurement of the  detached electrons:  If electrons of fixed energy emerge from a point source, and these electrons are accelerated by a uniform electric field toward a microchannelplate detector, then two trajectories go from the source to any given point on the detector \cite{Bracher1998a}.  Along each trajectory the wave function accumulates a phase proportional to the classical action of that path.  The resulting electron distribution at the detector shows a set of concentric interference fringes, associated with the two paths linking the source with the detection point.  More complicated paths and associated interference structures arise in photoionization experiments \cite{Kondratovich1984a,Kondratovich1984b,Kondratovich1990b,Bordas1998a}.  The observation and interpretation of such interference patterns is called photodetachment or photoionization microscopy. These techniques were pioneered by Blondel et al.\ \cite{Blondel1996a,Blondel1999a}, and Bordas, Vrakking and co-workers \cite{Nicole2002a,Bordas2003a,Lepine2004a}, respectively.  Quantum calculations \cite{Slonim1976a,Demkov1982a,Kramer2002a} yield excellent agreement with the experimental data, and in turn allow the determination of electron affinities from the interference patterns with unprecedented accuracy \cite{Blondel1999b,Blondel2005a}.

\begin{figure*}
\begin{center}
\includegraphics[width=\textwidth]{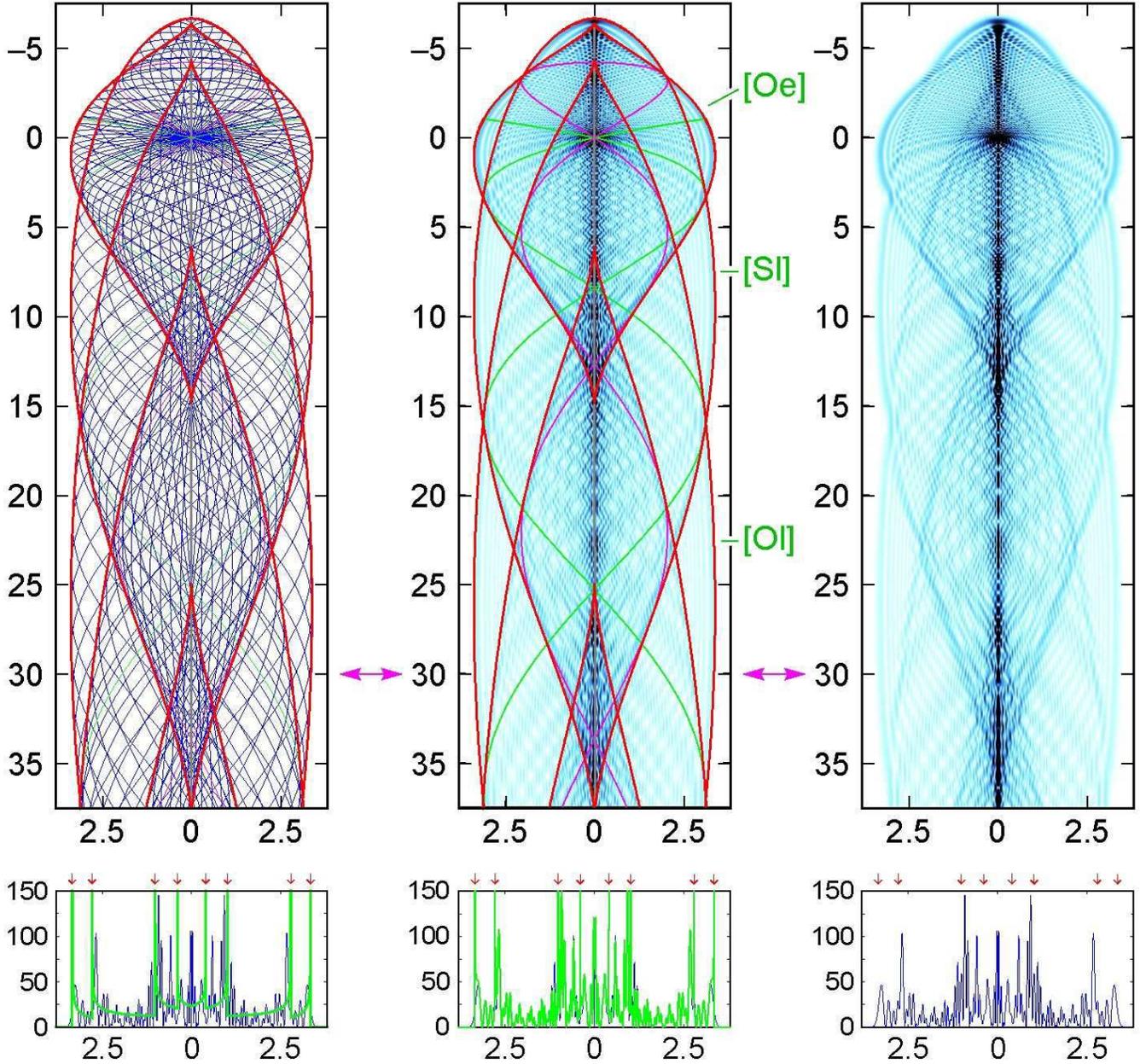}
\caption{
Motion of electrons emitted from a fixed energy, isotropic point source at the origin into parallel fields. (Parameters used: $E = 10^{-4}\,$eV, ${\cal E} = 15\,$V/m, ${\cal B} = 0.02\,$T, corresponding to a force parameter $\eta = v_0{\cal B} / {\cal E} = 7.908$ (\protect{\ref{eq:Class0.3}}).  All dimensions are in $\mu$m.)  The classical trajectory field (top left panel, blue curves) traces out an intersecting set of caustic surfaces (red curves) that separate regions with different numbers of classical paths connecting the source to a given destination point.  The maximum number (here, eight) occurs in the diamond-shaped entities along the symmetry axis.  The trajectory field contains two closed orbits that return to the source \protect{\cite{Peters1997a}}, a ``snake'' orbit (green), and a ``balloon'' orbit (purple). --- The center panel depicts the density profile of the electrons in primitive semiclassical approximation (blue plot), superimposed onto the caustic surfaces (red) and the closed orbits.  The caustics clearly delineate the features of the density distribution, and the interference pattern shows approximate symmetry with respect to the closed ``snake'' orbit.  (The letter combinations (green) refer to the classification of the caustics in Section~\protect{\ref{sec:Caustic}}.) --- The quantum density profile is displayed in the top right panel.  Its agreement with the semiclassical result (center panel) is remarkable.  --- Bottom row: Profiles of the radially integrated electronic density distribution $n(\hat\rho)$ at a distance $\hat z = 30\,\mu$m from the source (indicated by purple arrows above) for a normalized flux of one electron per second, in units of m$^{-2}$.  The green curve in the left panel denotes the purely classical cross section.  It is strongly modified by semiclassical interference (center panel). Both classical and semiclassical density profiles diverge at the positions of the caustics (marked by red arrows).  Otherwise, the semiclassical density profile quantitatively matches the quantum result (blue curve, right panel).
\label{fig:Primer0.1}}
\end{center}
\end{figure*}

In this paper we predict the distribution of electrons that would be seen in an ideal photodetachment microscope placed in parallel, uniform, static electric and magnetic fields $\boldsymbol{\cal E}$ and $\boldsymbol{\cal B}$.  With the idealization that the electrons come from a point source at fixed energy, we compute the spatial distribution of electrons, and predict their current density arriving at various points on a detector (Fig.~\ref{fig:Primer0.1}).

A weak cw laser field acting on a negative ion gives rise to outgoing, coherent, fixed-energy electron waves (compare a vibrator in a ripple tank).  The angular distribution of these waves is governed by selection rules for the initial and final state; for simplicity, we assume a spherically symmetric emission pattern \footnote{Near-threshold photodetachment into $s$--waves is the commonly observed case \cite{Blondel1996a,Blondel1999a,Blondel1999b}, with H$^-$ being an important exception \cite{Bryant1987a}.  A systematic way of dealing with multipole sources is developed in: C.~Bracher, T.~Kramer, and M.~Kleber, Phys.\ Rev.~A {\bf 67}, 043601 (2003).}.
The outward propagation of these waves can be described by considering the evolution of the corresponding radially outgoing classical trajectories.  As the paths bend in the applied electric and magnetic fields, the waves refract to follow them.   For weak external fields, the classical range of motion of the ejected electrons extends far beyond the residual atomic core, so their interaction with the core is negligible, and rescattering is limited to a small fraction of the initial outgoing wave.  It is a good approximation to neglect the presence of the atom after the photodetachment event altogether.  Then the dynamics of the electrons is governed solely by the externally applied fields.  (Rescattering is addressed e.~g.\ in Ref.~\cite{Fabrikant1989a,Fabrikant1994a,Fabrikant2002a}.)

In parallel fields aligned with the $z$ axis, every electron undergoes uniformly accelerated motion along the symmetry axis while orbiting on a circular track in the perpendicular $x-y$ plane (Section~\ref{sec:Classical}).  The initial emission angle $\theta'$ of the orbit relative to the field direction fixes both the energy contained in the motion along the $z$ axis and the energy available for transverse cyclotron motion, and these two energies are separately conserved.  The classical spatial distribution is determined by the set of all trajectories, combined with the angular characteristics of the source (isotropic here).  The upper left panel of Fig.~\ref{fig:Primer0.1} displays a profile of the resulting trajectory pattern (blue) for ${\cal E} = 15\,$V/m, ${\cal B} = 0.02\,$T, and $E = 10^{-4}\,$eV.  (The ordinate denotes the radial distance $\hat\rho$ from the $z$ axis.  All dimensions are in $\mu$m; axes are not to scale.)  The trajectories trace out an infinite sequence of \emph{caustic surfaces} (red curves); if we move a detector across a caustic surface, the number of classical paths arriving at the detector changes.  These caustic surfaces usually begin and end on the symmetry axis in a rotationally symmetric \emph{cusp}.  We will return to the caustic surfaces and their singularities in Section~\ref{sec:Caustic}.

The lower left plot of Fig.~\ref{fig:Primer0.1} shows a profile of the integrated classical electron density profile $n_{\rm cl}(\hat\rho) = 2\pi\hat\rho\cdot\rho_{\rm cl}(\hat\rho)$ (green), evaluated at a distance $\hat z = 30\,\mu$m from the source.  It is marked by divergences at the intersections with the caustic surfaces.  This classical result reproduces the overall trends in the quantum distribution (blue graph), but it lacks the oscillations evident in the quantum curve. They are produced by interference among waves traveling along various classical paths from the source to each point on the detector.  A \emph{primitive semiclassical} calculation (Section~\ref{sec:Semi}) \cite{Delos1986a,Schulman1981a} describes the interference (center column).  In this calculation the wave function at each point is a sum, in which every classical trajectory linking the source to the specified point contributes one term:  Its magnitude is the square root of the classical density associated with that trajectory and its neighbors, and its phase is given by the classical action (plus a correction $-\frac{\mu\pi}2$, where $\mu$ denotes the \emph{Maslov index} for that path).  An interesting feature of the semiclassical distribution is its approximate symmetry with respect to \emph{closed orbits} (green and purple curves).  Hence, their significance extends beyond their role in the total photodetachment cross section \cite{Peters1994a,Peters1997a}. 

A quantum calculation (Section~\ref{sec:Quantum}) that makes no reference to classical orbits is shown in the right column of Fig.~\ref{fig:Primer0.1}.  For a point source, the wave function is proportional to the \emph{energy Green function} $G(\mathbf{\hat r}; E)$ for an electron in the external potential.  This function can be expanded into a rapidly convergent series over Landau states, the eigenstates of the cyclotron motion \cite{Fabrikant1991a,Kramer2001a}.  This expansion yields both the total detachment cross section and the spatial electron distribution.  Although the mathematical structure of the quantum calculation bears little resemblance to the classical analysis, the semiclassical results approximate the quantum results remarkably well, as a comparison between the center and right panels in Fig.~\ref{fig:Primer0.1} reveals.  The only exception is at the caustic surfaces where the classical density diverges.  \emph{Uniform approximations} (not shown) correct the divergent behavior near caustics, and result in density profiles that are practically indistinguishable from the exact quantum result.  The very satisfying agreement of quantum and semiclassical results for this problem is further illustrated in Section~\ref{sec:Examples}.  Thus the features of the quantum density distribution are best explained from the classical trajectory fields.

The purpose of this paper is to present a complete analysis of the classical trajectories and their associated caustics, and to use that analysis to compute predictions of the features that might be seen in a photodetachment microscope in parallel electric and magnetic fields.  The equation of motion is trivial:  A superposition of ballistic motion in the $z$--direction, and cyclotron motion in the $x-y$ plane.  Complex behavior only arises when we consider the dynamics of the whole family of electron trajectories, most naturally formulated as an exercise in Hamilton--Jacobi theory.  The hardest part of the problem is understanding the structure of the caustics --- they are either delightfully or painfully complex, and a long analysis is required to describe them quantitatively.  Once this is done, it is easy to construct primitive semiclassical approximations for the wave function everywhere, and to derive uniform approximations that correct the divergences near the caustics for $\rho > 0$.

In addition, we show how to compute a quantum wave function for this system, and we demonstrate that the semiclassical approximations are usually in very good agreement with the quantum solution.

\section{Classical Mechanics}
\label{sec:Classical}

We first review the properties of the classical motion of an electron (charge $q = -e$, mass $m$) that is accelerated toward positive $\hat z$ in the fields $\boldsymbol{\cal E} = -{\cal E}\mathbf{e}_z$ and $\boldsymbol{\cal B} = -{\cal B}\mathbf{e}_z$, and choose as corresponding electromagnetic potentials $(\Phi,\mathbf A)$:
\begin{equation}
\label{eq:Class0.1}
\Phi(\mathbf{\hat r}) = \boldsymbol{\cal E} \hat z \;, \quad 
\mathbf A(\mathbf{\hat r}) = \frac12(\boldsymbol{\cal B} \times\mathbf{\hat r}) = \frac{{\cal B}}2 \bigl( \hat y, -\hat x, 0 \bigr)^T \;.
\end{equation}
(Physical coordinates $\mathbf{\hat r} = (\hat x,\hat y,\hat z)$ are marked with accents to distinguish them from their scaled dimensionless counterparts introduced below.  Primed symbols $\mathbf{\hat r'} = (\hat x',\hat y',\hat z')$ refer to the source rather than the destination.)  

The electrons are emitted from the source with fixed positive kinetic energy $E$, i.e., constant initial velocity $v_0 = \sqrt{2E / m}$, but arbitrary initial direction, defined by their polar angles $\theta'$ and $\phi'$.  (In the quantum case, tunneling sources with $E<0$ \cite{Bracher1998a} are also admissible.)  The spiraling motion of the electron in the magnetic field provides natural frequency and length scales in the problem, the Larmor frequency $\omega_L = e{\cal B} / 2m$ and the maximum cyclotron orbit diameter $d = v_0 / \omega_L$, so we introduce dimensionless time and position variables via:
\begin{equation}
\label{eq:Class0.2}
t = \omega_L \hat t \,,\quad x = \frac{\hat x}{d} \,,\quad y = \frac{\hat y}{d} \,,\quad 
z = \frac{\hat z}{d} \,,\quad \rho = \frac{\hat\rho}{d} \;.
\end{equation}
In this fashion, the dependence of the result on the parameters is eliminated except for one remaining variable, which we choose to be the ratio $\eta$ of the magnetic and electric forces in the system:
\begin{equation}
\label{eq:Class0.3}
\eta = v_0{\cal B} / {\cal E} \;.
\end{equation}

\subsection{Equations of motion}
\label{subsec:Class1}

We first characterize the trajectories in the problem.  The standard minimal coupling Lagrangian ${\cal L}(\mathbf{\hat r}, \rmd\mathbf{\hat r}/\rmd t)$ with the potentials (\ref{eq:Class0.1}) is
\begin{equation}
\label{eq:Class1.1}
\begin{split}
{\cal L} & = \frac m2 \left( \frac{\rmd\mathbf{\hat r}}{\rmd{\hat t}} \right)^2 
	- q\Phi(\mathbf{\hat r}) + q \frac{\rmd\mathbf{\hat r}}{\rmd\hat t} \cdot \mathbf A(\mathbf{\hat r}) \\
& = mv_0^2 \left[ \frac12 \bigl( \dot x^2 + \dot y^2 + \dot z^2 \bigr) 
	+ \frac2\eta z + x\dot y -\dot x y \right] \;,
\end{split}
\end{equation}
(where the dots indicate differentiation with respect to $t$).  We introduce a complex coordinate $\hat s = \hat\rho \rme^{\rmi\phi} = \hat x + \rmi\hat y$ and its scaled equivalent $s = \hat s/d$.  The equations of motion then read:
\begin{equation}
\label{eq:Class1.2}
\ddot s + 2\rmi\dot s = 0 \;,\quad \ddot z - 2/\eta = 0 \;,
\end{equation}
and their integration is straightforward.  For a particle with emission angles $(\theta',\phi')$, we find
\begin{equation}
\label{eq:Class1.3}
\begin{split}
s(t) & = \sin\theta'\sin t \rme^{\rmi(\phi' - t)} + s' \;, \\
z(t) & = t^2 / \eta + t\cos\theta' + z' \;.
\end{split}
\end{equation}
The second of these equations is the familiar uniform acceleration in $z$--direction, while the first describes the circular cyclotron orbits in the magnetic field.  If the source is located at the origin ($s' = 0$), the motion may be expressed in cylindrical coordinates $\rho$ and $\phi$ instead:
\begin{equation}
\label{eq:Class1.4}
\rho(t) = \sin\theta' \left| \sin t \right| \;,\quad
\phi(t) = \phi' - (t \mod\pi) \;.
\end{equation}
The polar angle $\phi(t)$ changes linearly during each cyclotron orbit cycle $k\pi < t < (k+1)\pi$, but discontinuously jumps by $\pi$ at each return to the $z$--axis ($t=k\pi$) \footnote{In contrast, the charge moves uniformly with an angular velocity $\omega_C = 2\omega_L$ if the \emph{center} of the orbit is chosen as the origin of the coordinate system.}.

Differentiation of Eq.~(\ref{eq:Class1.3}) yields the particle velocity:
\begin{equation}
\label{eq:Class1.5}
\dot s(t) = \sin\theta'\rme^{\rmi(\phi' - 2t)}\;,\quad
\dot z(t)  = 2t / \eta + \cos\theta' \;,
\end{equation}
Again, the polar angle of the velocity $\dot s(t)$ changes uniformly, albeit with twice the frequency.  Finally, we also state the (scaled) canonical momentum components $p_x = \partial{\cal L}/ \partial\dot x = v_0\hat p_x$ etc.\ connected with these trajectories.  From Eq.~(\ref{eq:Class1.1}), we find:
\begin{equation}
\label{eq:Class1.6}
\begin{split}
p_s(t) & = mv_0^2 \left( \dot s + \rmi s \right) 
	= mv_0^2 \bigl( \sin\theta'\cos t \rme^{\rmi(\phi' - t)} + \rmi s' \bigr) \;, \\
p_z(t) & = mv_0^2\dot z = mv_0^2 \left( 2t / \eta + \cos\theta' \right) \;.
\end{split}
\end{equation}
%
%

\subsection{Action functionals}
\label{subsec:Class2}

The equations of motion (\ref{eq:Class1.3}) describe the path of a charge as a function of its emission angles $(\theta',\phi')$ and initial velocity $v_0$, and the temporal variable $t$ parametrizes the trajectory.  However, in an experiment, the electron travels from a known initial point $\mathbf{\hat r'}$ to an observed final point $\mathbf{\hat r}$ on the detector.  Hamilton--Jacobi theory is especially suitable dealing with paths defined by such initial and final conditions, and an appropriate quantity to carry out calculations is the action integral ${\cal S}(\mathbf{\hat r}, \mathbf{\hat r'};\hat T)$ for a general source point $\mathbf r'$ and field point $\mathbf r$.

By definition, the action for a given path is the integral over the Lagrangian ${\cal L}(\mathbf{\hat r}, \rmd\mathbf{\hat r}/\rmd\hat t)$ (\ref{eq:Class1.1}).  Inserting the trajectory (\ref{eq:Class1.3}), we find initially:
\begin{equation}
\label{eq:Class2.1}
\begin{split}
{\cal S} & = \int_0^{\hat T} \rmd \hat t {\cal L}(\mathbf{\hat r}, \rmd\mathbf{\hat r}/\rmd \hat t) \\
& = \frac{mv_0^2}{2\omega_L} \Bigl( \frac{8 T^3}{3\eta^2} + \frac{4T(T\cos\theta' + z')}\eta 
+ T\cos^2\theta' \\
& \qquad\qquad + \sin T\cos T\sin^2\theta' + 2(x'y - xy') \Bigr) \;,
\end{split}
\end{equation}
where $T = \omega_L\hat T$.  Next, we must eliminate the emission angle $\theta'$ by demanding that the trajectory $\mathbf{\hat r}(\hat t)$ (\ref{eq:Class1.3}) passes through the destination $\mathbf{\hat r}$ after a time of flight $\hat T$---for instance, from Eq.~(\ref{eq:Class1.4}) we find $\sin^2\theta' = (s-s')^2 / \sin^2 T$.  The action then can be written as the sum of a longitudinal and a transverse part,
\begin{equation}
\label{eq:Class2.2}
{\cal S}(\mathbf r, \mathbf r'; T) = 
{\cal S}_\perp(s, s'; T) + {\cal S}_\|(z, z'; T) \;,
\end{equation}
which read respectively:
\begin{equation}
\label{eq:Class2.3}
{\cal S}_\perp(s, s'; T) = \frac{mv_0^2}{2\omega_L} \left( |s - s'|^2 \cot T + 2(x'y - xy') \right) ,
\end{equation}
\begin{equation}
\label{eq:Class2.4}
{\cal S}_\|(z, z'; T) = \frac{mv_0^2}{2\omega_L} \left( \frac{(z - z')^2}T + \frac{2(z + z')T}\eta - \frac{T^3}{3\eta^2} \right) .
\end{equation}
These results for the classical action in a two-dimensional magnetic field, and for the linear potential in one dimension, are well known \cite{Schulman1981a}.

Because ${\cal S}(\mathbf r,\mathbf r'; T)$ is a single-valued function, exactly one trajectory with given time of flight $T$ will link $\mathbf r$ and $\mathbf r'$.  Its initial momentum $\mathbf p' = - \boldsymbol\nabla'{\cal S}$ is given by:
\begin{equation}
\label{eq:Class2.5}
\begin{split}
p_s' & = -\omega_L \Bigl( \frac{\partial{\cal S}}{\partial x'} + 
\rmi\frac{\partial{\cal S}}{\partial y'} \Bigr)
= mv_0^2 \bigl[ (s - s') \cot T + \rmi s \bigr] \;, \\
p_z' & = -\omega_L \frac{\partial{\cal S}}{\partial z'} 
= mv_0^2 \Bigl[ \frac{z - z'}T - \frac T\eta \Bigr] \;. 
\end{split}
\end{equation}
We now assume $\mathbf r' = \mathbf 0$.  
The initial momentum along the $z$--axis is $p_z' = mv_0^2\cos\theta'$, so the emission angle as a function of $\mathbf r$ and $T$ follows from Eqs.~(\ref{eq:Class1.4})--(\ref{eq:Class1.6}):
\begin{equation}
\label{eq:Class2.6}
\begin{split}
\theta'(\mathbf r; T) &= \arccos(z/T - T/\eta) \;,\\
\phi'(\mathbf r; T) &= \phi + (T \mod \pi) \;,
\end{split}
\end{equation}
where $\phi$ is the polar angle of the destination $\mathbf r$.  Similarly, the final momentum $\mathbf p$ of the particle at its destination $\mathbf r$ is given by the derivative $\boldsymbol\nabla{\cal S}$:
\begin{equation}
\label{eq:Class2.7}
\begin{split}
p_s & = \omega_L \Bigl( \frac{\partial{\cal S}}{\partial x} + 
\rmi\frac{\partial{\cal S}}{\partial y} \Bigr)
= mv_0^2 \bigl[ (s - s') \cot T + \rmi s' \bigr] \;, \\
p_z & = \omega_L \frac{\partial{\cal S}}{\partial z} 
= mv_0^2 \Bigl[ \frac{z - z'}T + \frac T\eta \Bigr] \;. 
\end{split}
\end{equation}
In particular, for $\mathbf r' = \mathbf 0$ the velocity component in $z$--direction at $\mathbf r$ (\ref{eq:Class1.6}) reads:
\begin{equation}
\label{eq:Class2.8}
\dot z(\mathbf r; T) = z/T + T/\eta \;.
\end{equation}

We define $E(\mathbf{\hat r},\mathbf{\hat r}';\hat T)$ as the energy of the electron that goes from $\mathbf{\hat r}'$ to $\mathbf{\hat r}$ in time $\hat T$.  Similarly, $E(\mathbf r; T)$ is the energy of a particle that starts from the origin ($\mathbf{\hat r}' = \mathbf 0$) and arrives at the scaled position $\mathbf r$ at the scaled time $T$.  These energy expressions follow from the derivative of the classical action:  $E(\mathbf{\hat r},\mathbf{\hat r}';\hat T) = -\partial{\cal S}(\mathbf{\hat r}, \mathbf{\hat r}';\hat T)/\partial\hat T$, so Eqs.~(\ref{eq:Class2.2})--(\ref{eq:Class2.4}) yield the decomposition:
\begin{equation}
\label{eq:Class2.9}
E(\mathbf r; T) = E_\perp(\rho; T) + E_\|(z; T) \;,
\end{equation}
with
\begin{gather}
\label{eq:Class2.10}
E_\perp(\rho; T) = \frac{mv_0^2}2 \frac{\rho^2}{\sin^2 T} 
                 \equiv \frac{mv_0^2}2 \,\epsilon_\perp(T) \;, \\
\label{eq:Class2.11}
E_\|(z; T) = \frac{mv_0^2}2 \Bigl( \frac zT - \frac T\eta \Bigr)^2 
           \equiv \frac{mv_0^2}2 \,\epsilon_\|(T) \;.
\end{gather}

The experiment provides a stationary source that emits particles with fixed energy $E = mv_0^2 / 2$, rather than with a definite time of flight $\hat T$.  Therefore, only those classical trajectories fulfilling the condition $E = E(\mathbf r; T)$ are acceptable solutions.  Hence, we may interpret Eq.~(\ref{eq:Class2.9}) as an implicit equation for the time of flight $T(\mathbf r; E)$ as a function of the particle energy:
\begin{equation}
\label{eq:Class2.12}
\epsilon(T) \equiv \frac{E(\mathbf r;T)}E = 
\frac{\rho^2}{\sin^2 T} + \Bigl( \frac zT - \frac T\eta \Bigr)^2 = 1 \;.
\end{equation}
While this transcendental equation must be solved numerically for $T$, this is the only instance in the classical description of motion in parallel fields that we have to resort to computation.  Every other quantity of relevance can be cast as a closed-form expression involving the time of flight $T$.  For example, the reduced action (Hamilton's characteristic function) ${\cal W}(\mathbf{\hat r}; \hat E)$ that governs the semiclassical phase follows from Eq.~(\ref{eq:Class2.12}) together with the time-dependent action (\ref{eq:Class2.2}) via:
\begin{equation}
\label{eq:Class2.13}
{\cal W}(\mathbf{\hat r}; E) = \int_{\mathbf 0}^{\mathbf{\hat r}} \mathbf{\hat p}\cdot\rmd\mathbf{\hat q} =
{\cal S}\bigl(\mathbf{\hat r}; \hat T(\mathbf{\hat r};E)\bigr) + E \hat T(\mathbf{\hat r};E) \;.
\end{equation}
In general, Eq.~(\ref{eq:Class2.12}) possesses a variety of solutions $T$.  Correspondingly, the reduced action ${\cal W}(\mathbf r; E)$, like other quantities of interest in the stationary theory, is a multi-valued function and depends on the specific trajectory chosen.
 
\subsection{Number of trajectories}
\label{subsec:Class3}

Clearly, the multiplicity of real solutions $T$ to the implicit equation (\ref{eq:Class2.12}) determines the number of classical trajectories with given energy $E$ that connect $\mathbf r' = \mathbf 0$ with $\mathbf r$ in parallel fields.  The semiclassical analysis of the problem rests on our knowledge of these trajectories.  Here, we establish some general properties of the set of solutions.  \footnote{The conjugate complex roots of Eq.~(\ref{eq:Class2.12}) formally correspond to ``tunneling trajectories'' in complex space-time.  Because they describe evanescent waves in the classically forbidden sector of motion, they play an important role in refining the semiclassical solution near caustic surfaces.  For a detailed analysis of the tunneling trajectories in a purely electric field, see Ref.~\protect{\cite{Bracher1998a}}.}

\begin{figure}
\begin{center}
\includegraphics[width=\columnwidth]{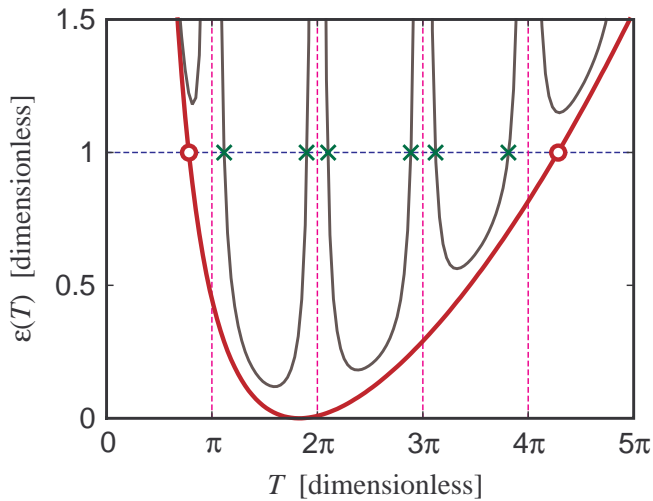}
\caption{Graphical solution for the number of trajectories from source to detector.  The heavy (red) curve is $\epsilon_\|(T)$, Eq.~(\protect{\ref{eq:Class2.10}}).  Its intersections with the line $\epsilon = 1$ define $T_-$ and $T_+$ (circles), the times of flight of electrons respectively emitted in the direction, and against the direction, of the electric force.  The light (black) curves trace $\epsilon(T)$, Eq.~(\protect{\ref{eq:Class2.12}}).  Each goes to infinity at $T = k\pi$, i.e., whenever a cyclotron orbit is complete.  Every intersection of these curves with $\epsilon = 1$ (crosses) represents the time of flight of a trajectory from the source to the specified point $\mathbf r$ on the detector.  There are at most two solutions within each cyclotron period $(k-1)\pi < T < k\pi$.  (In this graph, $\eta=11$, $z = 3$, and $\rho = 0.3$.)
\label{fig:Class3.1}}
\end{center}
\end{figure}

To learn more about Eq.~(\ref{eq:Class2.12}) is is useful to sketch the scaled particle energy $\epsilon(T)$ as a function of the time of flight $T$ (Fig.~\ref{fig:Class3.1}).  In the graph, we plot the longitudinal contribution to the energy $\epsilon_\|(T)$ (\ref{eq:Class2.11}) (red line).  It approaches infinity both for $T \rightarrow 0$ and $T \rightarrow \infty$, runs for $z > 0$ smoothly through a minimum $\epsilon(T_{\rm ff}) = 0$ at $T_{\rm ff} = \sqrt{\eta z}$ which corresponds to the ``freely falling'' motion of a charge starting at rest, and intersects the line $\epsilon_\|(T_\pm) = 1$ at the scaled times $T_\pm$:
\begin{equation}
\label{eq:Class3.1}
T_\pm = \frac12 \Bigl| \sqrt{\eta^2 +4\eta z} \,\pm \eta  \Bigr| \;.
\end{equation}
These are the flight times of electrons emitted parallel to the fields.  For $z \geq 0$ their difference $\Delta T = \eta$ is constant and equal to the return time for the parallel closed orbit.  (Note that in physical coordinates, $\hat T_\pm = T_\pm / \omega_L$ does not depend on the magnetic field.)  Since $\epsilon(T) \geq \epsilon_\|(T)$, all real solutions of Eq.~(\ref{eq:Class2.12}) are located in the interval $T_- \leq T \leq T_+$.

Adding the perpendicular component $\epsilon_\perp(T)$, Eq.~(\ref{eq:Class2.10}), yields the full dependence of $\epsilon(T)$ on $T$ (black curves in Fig.~\ref{fig:Class3.1}).  For $\rho > 0$, this ``magnetic'' part of the energy diverges whenever the time of flight $T_k = k\pi$ $(k=1,2,3,\ldots)$, is a multiple of the cyclotron orbit time, when every trajectory must return to the $z$--axis.  Halfway between these singularities, $\epsilon_\perp(T)$ adopts its minimum value $\epsilon_\perp\bigl[(k - \frac12)\pi\bigr] = \rho^2$.  For $\rho > 1$, no real solutions to Eq.~(\ref{eq:Class2.12}) exist, and the sector of classically allowed motion is entirely contained inside (but not filling) the cylinder $\rho \leq 1$.  It is a simple exercise to verify that within each cyclotron orbit cycle $(k-1)\pi < T < k\pi$, $\epsilon(T)$ is a convex function that has a single minimum $\epsilon_{\min}^{(k)}$.  Its value determines the number of trajectories in this cycle:  For $\epsilon_{\min}^{(k)} < 1$ there exist exactly two times of flight $T_<^{(k)}$ and $T_>^{(k)}$ that correspond to an \emph{early} and a \emph{late} trajectory that both complete $k-1$ cyclotron orbits before arriving at $\mathbf r$ (marked by $\times$ in Fig.~\ref{fig:Class3.1}).  

With increasing radial distance $\rho$, $\epsilon_{\min}^{(k)}$ grows until it becomes equal to unity, causing the two trajectories to coalesce into a single solution.  For even greater distances $\rho$, $\epsilon_{\min}^{(k)} > 1$ holds, and no classical trajectories exist in the cycle.  (This transition spawns a complex conjugate pair of solutions that make up a \emph{tunneling trajectory}.)  We infer that with increasing radius $\rho$, the number of solutions $N(\rho,z)$ successively drops by two.  (See Fig.~\ref{fig:Primer0.1}.)

Consequently, for any given distance $z$, the maximum number of trajectories occurs in the vicinity of the $z$--axis ($\rho\rightarrow 0$).  For infinitesimal $\rho$, the lateral energy $\epsilon_\perp(T)$ becomes negligible, except in the immediate neighborhood of the return times $T_k$, where $\epsilon_\perp(T)$ still diverges.  As a result, the late and early trajectories $T_>^{(k-1)}$, $T_<^{(k)}$ in consecutive cycles generally tend towards the singularities at $T_k=k\pi$, while the fastest and the slowest solutions approach the parallel times $T_\pm$ (\ref{eq:Class3.1}), respectively:  For $\rho \rightarrow 0$, every cyclotron orbit period $(k-1)\pi < T < k\pi$ that intersects the interval $T_- \leq T \leq T_+$ contributes two real solutions to Eq.~(\ref{eq:Class2.12}).  Thus, the total number of solutions $N(\rho\rightarrow 0, z)$ depends only on the ratio of the interval sizes, and their relative positions.  It is easy to see that for $z > 0$,
\begin{equation}
\label{eq:Class3.2}
N(\rho\rightarrow 0, z) = 2 \bigl[ \lfloor \eta/\pi \rfloor + n(z) \bigr] \;,
\end{equation}
where $\lfloor \ldots \rfloor$ denotes the integer part, and $n(z)$ is either 1 or 2, depending on the distance $z$.  (In particular, for $T_{\rm ff} = k\pi$, $n(z) = 2$ holds, while $n(z) = 1$ for $T_{\rm ff} = (k - \frac12)\pi$.)  We infer that the ratio of forces $\eta$ presents a convenient control parameter that adjusts the degeneracy of the solution set.  We also note that no matter how small we choose the magnetic field ${\cal B}$, in the vicinity of the ``focal points'' $z = k^2\pi^2 / \eta$ always a region with at least four coexisting trajectories exists \cite{Kramer2001a}.  Figure~\ref{fig:Primer0.1} shows an example with up to eight interfering trajectories.

For a destination on the $z$--axis ($\rho =0$), the solutions $T_>^{(k-1)}$ and $T_<^{(k)}$ coincide with the orbit time $T_k = k\pi$.  In this case, the polar angle of the destination $\phi$ is no longer defined, and the emission angle $\phi'$ (\ref{eq:Class2.6}) is unrestricted:  An entire cone of trajectories connects the origin with $\mathbf r$, reflecting the cylindrical symmetry of the situation.  In particular, this covers the situation $\mathbf r = \mathbf 0$, where the theory yields the orbits returning to the source:  Besides the ``uphill'' closed orbit emitted antiparallel to the electric force, having time of flight $\Delta T = \eta$ (\ref{eq:Class3.1}) and action ${\cal W}_{\rm el}(\mathbf 0; E) = 2 \eta E / (3\omega_L)$ (\ref{eq:Class2.13}), there exist $\lfloor \eta/ \pi \rfloor$ additional ``magnetic'' closed orbits with time of flight $T_k = k\pi$, emission angle $\theta'_k = \arccos(- k\pi/\eta)$, and action:
\begin{equation}
\label{eq:Class3.3}
{\cal W}_{\rm magn}^{(k)}(\mathbf 0; E) = \frac{k\pi E}{\omega_L} \Bigl( 1 - \frac{k^2\pi^2}{3\eta^2} \Bigr)
\;,
\end{equation}
where $k < \eta/\pi$.  In ($\rho,z$) space, these orbits have the shape of either ``balloons'' (even $k$) or ``snakes'' (odd $k$).  For the latter, both $\dot\rho$ (\ref{eq:Class1.4}) and $\dot z$ (\ref{eq:Class2.8}) are zero at $t = T_k/2$:  The electron stops, turns around, and retraces its path back to the source.  (See also Fig.~\ref{fig:Primer0.1}.) A semiclassical study of the total photodetachment cross section in parallel fields has been based on these returning trajectories \cite{Peters1994a,Peters1997a}.

\section{The Caustic Surface}
\label{sec:Caustic}

Placing the detector at fixed distance $z$ from the source, as we move outward in $\rho$, at certain discrete distances the number of classical trajectories arriving at the point $(\rho, z)$ drops.  These locations, at which an infinitesimal shift changes the number of solutions, make up the boundaries of classically allowed motion, the caustic surfaces \cite{Berry1981a}.  A simple way of illustrating the caustics is depicted in Fig.~\ref{fig:Primer0.1} (top left panel):  Using the equation of motion (\ref{eq:Class1.3}), (\ref{eq:Class1.4}), we draw a series of classical trajectories with increasing emission angle $\theta'$ (blue curves).  These trajectories then trace out their bounding surfaces (red lines).  The result is a structure of considerable complexity.  (Similar presentations, for varying values of the ratio of forces $\eta$ (\ref{eq:Class0.3}), can be found in Ref.~\cite{Peters1994a}.)

We would like to obtain a closed functional form $\rho(z)$ representing the caustics, but no such explicit dependence can be found.  Instead, we employ a \emph{temporal parametrization} of the caustic surfaces $\bigl[\rho(t), z(t)\bigr]$ which yields the location of a point on the caustic as a function of the time of flight $t$ required to reach it from the source.
In this parametrization, each cyclotron period $(k-1)\pi < t < k\pi$ gives rise to an individual caustic surface.  Additional \emph{focal line} segments on the symmetry axis (where trajectories cross through $\rho = 0$ after $k$ complete cyclotron periods, i.e., $t = k\pi$) also may be regarded as caustics.

The resulting caustic surfaces themselves have a surprising degree of complexity:
Seven different types occur (Fig.~\ref{fig:Caustic4.1}) that regularly change into each other with increasing parameter $\eta$.  This evolution of the caustic structure is illustrated in Fig.~\ref{fig:Caustic5.1}.  These results follow from an elementary but lengthy argument based on the irregular points of the parametrization (that may or may not show up as singularities in the caustic curve in space $\rho(z)$).  We only sketch it here; a complete proof is given in the appendix.  The reader who is not interested in mathematical detail may simply scrutinize Figs.~\ref{fig:Caustic4.1} and~\ref{fig:Caustic5.1} together with Table~\ref{tab:Caustic5.1}, and then skip to Section~\ref{sec:Semi}.

\subsection{Temporal parametrization of caustics}
\label{subsec:Caustic1}

A point on the caustic surface $\mathbf r$ is characterized by two coalescent trajectories connecting it to the source, which we choose to be the origin.  In the time-of-flight picture, this requires not only that the energy $E(\mathbf r; t)$ (\ref{eq:Class2.9}) for the trajectory matches the particle energy ($\epsilon(t) = 1$), but that additionally $E(\mathbf r; t)$ as a function of $t$ attains a minimum there (two roots of $\epsilon(t) = 1$ coalesce).  Hence, $\mathbf{\hat r}$ will be located on the caustic surface if:
\begin{equation}
\label{eq:Caustic1.1}
\frac{\partial S}{\partial{\hat t}}(\mathbf{\hat r}; \hat t) = - E \quad \mathrm{and} \quad
\frac{\partial^2 S}{\partial\hat t^2}(\mathbf{\hat r}; \hat t) = 0 
\end{equation}
hold simultaneously, or equivalently,
\begin{equation}
\label{eq:Caustic1.1a}
\epsilon(t) = 1 \quad \mathrm{and} \quad \frac{\partial\epsilon}{\partial t} = 0 \;.
\end{equation}
We already examined the first requirement, Eq.~(\ref{eq:Class2.12}), in Section~\ref{subsec:Class3}.  From Eqs.~(\ref{eq:Class2.2})--(\ref{eq:Class2.4}), the second condition reads in explicit form:
\begin{equation}
\label{eq:Caustic1.2}
-\frac12 \, \frac{\partial\epsilon}{\partial t} = \frac{\rho^2\cos t}{\sin^3 t} + \frac{z^2}{t^3} - \frac t{\eta^2} = 0 \;. 
\end{equation}

The set of Eqs.~(\ref{eq:Class2.12}) and (\ref{eq:Caustic1.2}) for the three variables $\rho$, $z$, and the time of flight $t$ implicitly characterize the caustic surfaces in the parallel field problem.  It would be nice if we could eliminate $t$ between these equations, and represent the caustics more directly as $\rho_{\mathrm {caustic}}(z)$.  (Indeed, this strategy works for a uniform pure electric field; one finds that the caustic is a paraboloid with the source as its focus \cite{Bracher1998a}.)  However, for nonzero magnetic field, the trigonometric quantities in Eqs.~(\ref{eq:Class2.12}) and (\ref{eq:Caustic1.2}) prevent elimination of $t$.

Instead, we can get a parametric representation of the caustics by the following method.  Using Eq.~(\ref{eq:Caustic1.2}), we express $\rho^2$ as a function of $z$ and $t$, and then insert the result into Eq.~(\ref{eq:Class2.12}); this yields a quadratic equation relating $z$ and $t$.  We put the solution to this equation back into Eq.~(\ref{eq:Caustic1.2}), and obtain a temporal parametrization of the caustic surfaces:
\begin{equation}
\label{eq:Caustic1.3}
\begin{split}
z_\pm(t) & = \frac{t^2 \bigl( t \pm \sqrt{A(\tau,t)} \bigr)}{\eta(t - \tau)} \;, \\
\rho_\pm(t) & = \frac{|\sin t|}\eta 
\sqrt{t\tau \Biggl[ 1 - \Biggl( \frac{t \pm \sqrt{A(\tau,t)}}{t - \tau} \Biggr)^2 \Biggr]} \;.
\end{split}
\end{equation}
where:
\begin{equation}
\label{eq:Caustic1.3a}
\tau = \tan t \;, \quad A(\tau, t) = \tau^2 - \frac{\eta^2\tau}t + \eta^2 \;.
\end{equation}
Signs must match in the pair of equations (\ref{eq:Caustic1.3}).  The corresponding emission angle follows from Eq.~(\ref{eq:Class2.6}):
\begin{equation}
\label{eq:Caustic1.4}
\cos\theta'_\pm(t) = \frac{t \Bigl(\tau \pm \sqrt{A(\tau,t)}\Bigr)}{\eta(t - \tau)} \;.
\end{equation}

\subsection{Existence of solutions}
\label{subsec:Caustic2}

The expression (\ref{eq:Caustic1.3}) shows that for a given time of flight $t$, there exist at most two corresponding points $\bigl[ \rho_\pm(t), z_\pm(t)\bigr]$ on the caustic surface.  These solutions are not necessarily real, however.  In order to qualify as a physical solution, a pair of inequalities must be fulfilled,
\begin{subequations}
\begin{equation}
\label{eq:Caustic2.2a}
A(\tau, t) \geq 0 \mathit{\ and\ } (t-\tau)^2 \geq \bigl(t \pm \sqrt{A(\tau, t)} \bigr)^2  
\mathit{\ \ for\ }\tau > 0 \;,
\end{equation}
\textsl{or}
\begin{equation}
\label{eq:Caustic2.2b}
A(\tau, t) \geq 0 \mathit{\ and\ } (t-\tau)^2 \leq \bigl(t \pm \sqrt{A(\tau, t)} \bigr)^2 
\mathit{\ \ for\ }\tau < 0 \;.
\end{equation}
\end{subequations}
The signs ``$\pm$'' refer to the branches in Eq.~(\ref{eq:Caustic1.3}).  Note that the ordering in the second inequality depends on the sign of $\tau = \tan t$.  (For $\tau=0$, both solutions exist and lie on the symmetry axis $\rho = 0$.)
\begin{figure}
\begin{center}
\includegraphics[width=0.75\columnwidth]{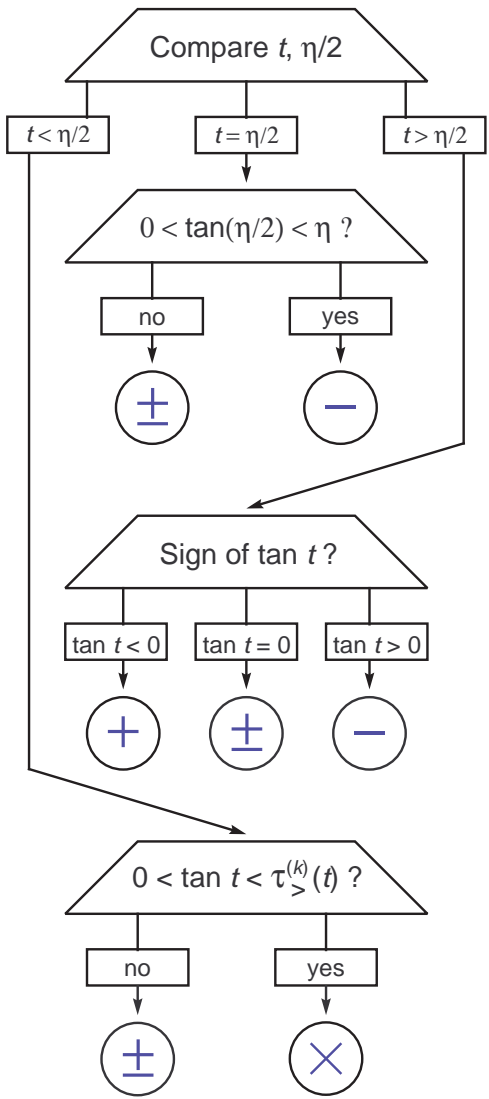}
\caption{An algorithm detailing the existence of real solutions in Eq.~(\protect{\ref{eq:Caustic1.3}}).
\label{flowchart}}
\end{center}
\end{figure}

An algorithm for selecting the upper $(+)$ and/or lower $(-)$ sign in Eqs.~(\ref{eq:Caustic1.3})--(\ref{eq:Caustic1.4}) consistent with these conditions is given in Fig.~\ref{flowchart}.  To understand it, it is best to think of $A(\tau, t)$ (\ref{eq:Caustic1.3a}) as a function of two independent variables.  For $t \geq \frac\eta2$, $A(\tau, t)$ is never negative, but for $t < \frac\eta2$, $A(\tau, t)$ usually has two roots in each cyclotron period $(k-1)\pi < t < k\pi$, the larger of which is:  
\begin{equation}
\label{eq:Caustic2.2c}
\tau_>^{(k)}(t) = \frac{\eta^2}{2t} \left(1 + \sqrt{1 - \frac{4t^2}{\eta^2}} \right) \;.
\end{equation}
(The period containing $t = \frac\eta2$ is the only possible exception.)  
We call the solutions of this transcendental equation the \emph{minimal times}:
\begin{equation}
\label{eq:Caustic2.2d}
\tau_>^{(k)} = \tan t_{\min}^{(k)} \;.
\end{equation}
With this information, the algorithm is easy to apply.  A proof is given in the appendix.  

As $t$ increases from $0$ to $\frac\eta2$, $\tau=\tan t$ sweeps from 0 toward infinity, and then from $-\infty$ to $+\infty$ an integer number of times, finally ending up (usually) at some finite value of $\tau$.  In this range of $t$, whenever $\tau$ lies between 0 and $\tau_>^{(k)}$, both ($+$) and ($-$) solutions are complex.  Two real solutions are created at $\tau_>^{(k)}$, and they persist while $\tau$ increases to $+\infty$, jumps back to $-\infty$ and increases again, until $\tau$ passes through zero, at which point they disappear (Fig.~\ref{flowchart}).

Once $t$ exceeds $\frac\eta2$, \textit{either} the ($+$) \textit{or} the ($-$) solution is real, according to whether $\tau<0$ or $\tau>0$, respectively, so $\rho(t)$, $z(t)$ are single-valued functions of $t$.  ($+$) and ($-$) solutions join each other smoothly when $\tau$ passes through infinity. 

\subsection{Irregular points on caustics}
\label{subsec:Caustic3}

Usually, the parametrization $\bigl[ \rho_\pm(t), z_\pm(t) \bigr]$ (\ref{eq:Caustic1.3}) is a smooth function of the time of flight $t$, and the resulting caustic surface in space is locally smooth as well.  We call such points on the caustic \emph{regular}.  For some isolated values $t_0$, however, one or both of the functions $\rho_\pm(t)$, $z_\pm(t)$ become singular.  While such \emph{irregular} points $(t_0,z_0,\rho_0)$ are rare, they are important for our understanding of the caustic structure because qualitative changes of the parametrization may only occur at these points.  Likewise, singularities of the caustic surfaces themselves are always connected to irregular points.

Because they are involved with changes in the number or the branch of solutions, the irregular points of the parametrization can be read off the flowchart algorithm in Fig.~\ref{flowchart}.  We identify eight different types:  (Case designations refer to the corresponding section in the appendix; the symbols are used in the figures.)

\paragraph{Maxima of $\rho(t)$ $(\Box)$.}
The irregular points are at $t_0 = (k - \frac12)\pi$, where $\tau = \tan t$ diverges and changes sign.  Here, the ($+$) and ($-$) solutions switch seamlessly into each other.  On the caustic surface, the corresponding irregular points mark the maxima of $\rho$ (case 2a), as well as the endpoints of snake-type closed orbits (case 2b).  The latter exist only for  $t < \frac\eta2$.

\paragraph{Top of caustic $(\boldsymbol\star)$.}
This irregular point is at $(t_0, z_0, \rho_0) = (\frac\eta2, -\frac\eta4, 0)$, the uphill turning point (case 2c).  There, the radicand in $\rho_\pm(t)$ vanishes for either the ($+$) or the ($-$) solution (the remaining solution is unaffected), and the number of physical solutions changes by one.  The caustic surface itself is usually smooth.

\paragraph{Minimal times $(\Diamond)$.}
Here, $t_0 = t_{\min}^{(k)}$ (\ref{eq:Caustic2.2d}) is a root of $A(\tan t,t) = 0$ (case 2d).  The singularity occurs only for $t < \frac\eta2$ and indicates a minimum of the time of flight from the source to the caustic.  At $t=t_0$, the solution pair $\bigl[ \rho_\pm(t), z_\pm(t) \bigr]$ springs into existence.

\paragraph{Complete cyclotron orbits $(\circ)$.}
For $t_0 = k\pi$ the caustics intersect the symmetry axis ($\rho_0 = 0$), and $\tan t$ changes sign (case 2e).  Caustic surfaces ``begin'' and ``end'' here.  In contrast to the previous cases, the irregular points at $z_0 = \frac{k\pi}\eta (k\pi \pm \eta)$ mark singularities in the caustic profile $\rho(z)$.  Usually, the caustic surface locally has the shape of a rotationally symmetric \emph{cylindrical cusp} \cite{Peters1997b}.  \emph{Focal line segments}, parts of the caustic structure where the trajectories are focused back onto the symmetry axis, connect the cusp singularities.

These irregular points are ``generic,'' i.e., they occur for almost all choices of the parameter $\eta$.  At some isolated values of $\eta$, however, the top of the caustic ($t_0 = \frac\eta2$) coincides with another irregular point, and the ensuing singularity has distinct properties.  Such values of $\eta$ are important when we try to figure out how the pattern of caustics evolves with increasing $\eta$.  Three cases arise:

\paragraph{Hyperbolic umbilic points $(\boldsymbol\times)$.}
When $\eta = 2k\pi$ and $t_0 = k\pi$ (case 3a), the time to reach the top of the caustic is a multiple of the cyclotron period, and the uphill turning point is shared by two locally degenerate caustic surfaces in the shape of a \emph{cylindrical cone} with opening angle $\frac\pi4$.  They are manifestations of a higher-order singularity known as \emph{hyperbolic umbilic} points \cite{Poston1978a}.  Also, a ``balloon''--type closed orbit is created as $\eta$ increases through $2k\pi$ \cite{Peters1994a}.

\paragraph{Bifurcation points $(\bigtriangledown)$.}
Here, $t_0 = t_{\min}^{(k)} = \frac\eta2$ holds for the minimal time (case 3b).  Then, the force parameter $\eta$ takes on one of its \emph{critical values} $\eta_{\rm crit}^k$ given by the roots of the transcendental equation:
\begin{equation}
\label{eq:Caustic3.0a}
\eta_{\rm crit}^{k} = \tan \bigl(\eta_{\rm crit}^{k} / 2 \bigr) \;,
\end{equation}
where $k = 1,2,\ldots$ refers to the interval $2(k-1)\pi < \eta_{\rm crit}^{k} < (2k-1)\pi$.  At these points, a bifurcation occurs that affects the temporal ordering of the uppermost caustic, but not its shape.

\paragraph{Creation of a closed orbit $\bigtriangleup$.}
Finally, for $\eta = (2k-1)\pi$ (case 3c), the time of flight for the antiparallel ``uphill'' closed orbit is an odd multiple of the cyclotron period.  At this value of $\eta$, a new ``snake''--type closed orbit is created \cite{Peters1994a}.

To obtain these properties, we studied the local behavior of the parametrization (\ref{eq:Caustic1.3}) near the irregular points.  Details are discussed in the appendix.

\subsection{Classification of caustics}
\label{subsec:Caustic4}

\begin{figure*}
\begin{center}
\includegraphics[width=\textwidth]{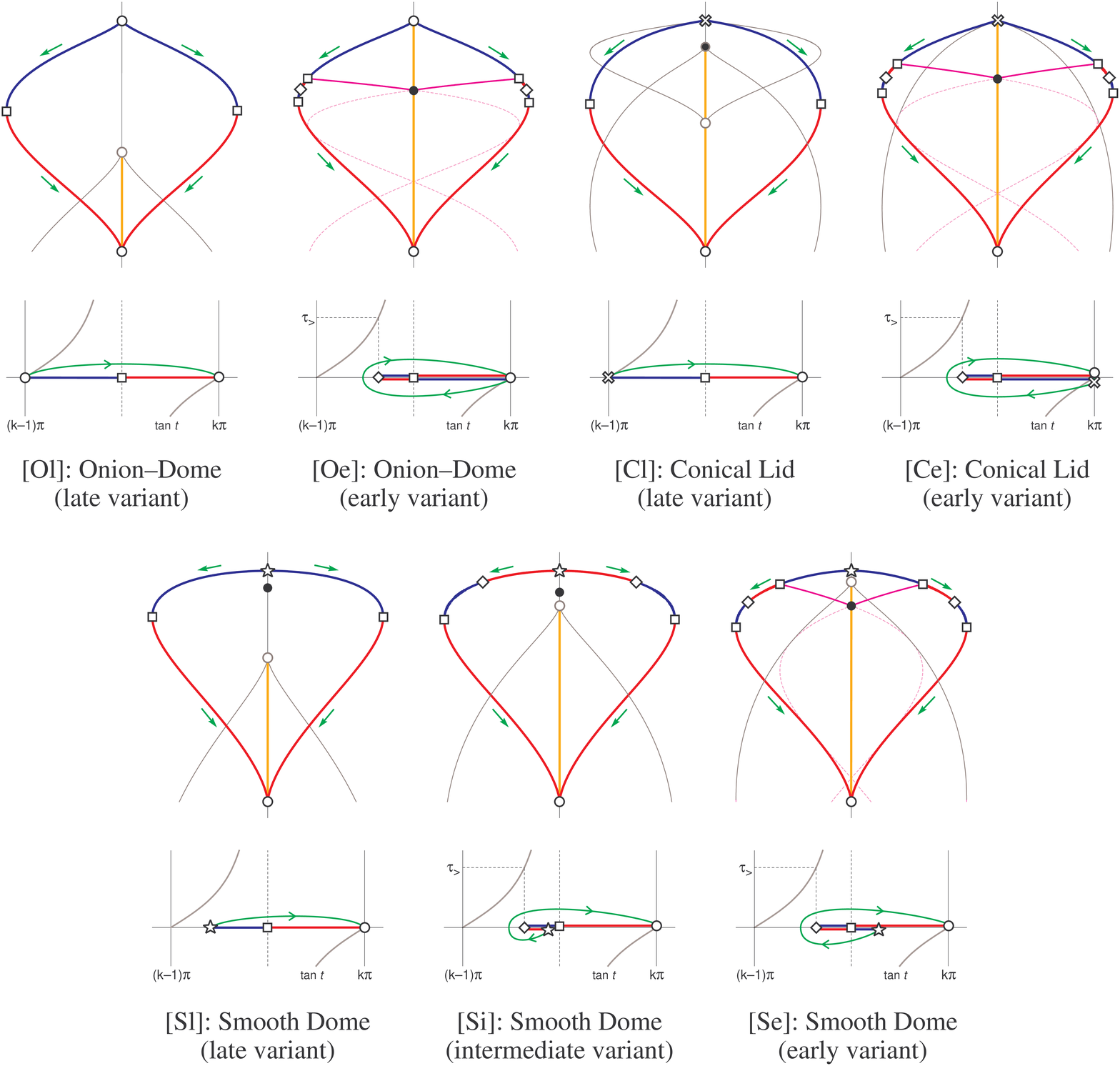}
\end{center}
\caption{The seven types of caustic surfaces occurring in the parallel field problem. They form three families with distinctive shapes of the turning surface: \emph{Onion-domes} (top left pair) come in \emph{late} (or no-closed-orbit) variants (left, $\eta < 2(k-1)\pi$) and \emph{early} (or closed-orbit) variants (right, $\eta > 2k\pi$).  The related late and early \emph{conical lids} (top right pair) are encountered for $\eta=2(k-1)\pi$ and $\eta=2k\pi$, respectively.  The bottom row depicts the \emph{smooth domes} (or \emph{inverted teardrops}) occurring for $2(k-1)\pi < \eta < 2k\pi$.  The three subtypes differ only in their parametrization, as explained in the text. ($k$ denotes the index of the caustic; i.e., electrons arrive at the caustic in their $k$th cyclotron orbit, $(k-1)\pi \leq t \leq k\pi$.) --- Beneath the picture of each caustic is a time axis showing the temporal parametrization (\protect{\ref{eq:Caustic1.3}}).  As $t$ follows the green curve and we switch between the ($+$) solution (red) and the ($-$) solution (blue), the path in space traces the caustic from the highest point to the lower cusp (green arrows).  Orange parts of the symmetry axis represent \emph{focal lines} of infinite degeneracy ($t=k\pi$), and purple curves are closed orbits. Symbols represent irregular points of the parametrization (see Appendix): Half-integer orbits ($\Box$, cases 2a, 2b), top of caustic ($\boldsymbol\star$, case 2c), the minimal times $t_{\rm min}^{(k)}$ ($\Diamond$, case 2d), cylindrical cusps ($\circ$, case 2e), and umbilical points ($\boldsymbol\times$, case 3a).  Black dots ($\bullet$) mark the source location. --- Parameters used (left to right): $\eta=1.5$, $k=2$; $\eta=7.908$, $k=1$ (cf.\ Fig.~\protect{\ref{fig:Primer0.1}});  $\eta=2\pi$, $k=2$ and $k=1$ (cf.\ Fig.~\protect{\ref{fig:Caustic3.1}}) (top row); $\eta=1.5$, $\eta=2.75$, $\eta=4$, all $k=1$ (bottom row).  (Force directed downward, caustics not to scale.)
\label{fig:Caustic4.1}
}
\end{figure*}

We listed eight different types of irregular points in the parametrization of the caustics.  However, when we look at the caustic surfaces in $(\rho,z)$--space, we find, happily, only two shapes of caustics that are present for general values of $\eta$, and a third occurring when $\eta$ is a multiple of $2\pi$.  They all resemble Christmas tree ornaments cusped at their bottom, but differing at their tops:  The uppermost region may be a cylindrical cusp shaped like an onion--dome (type [O]; the slope ${\rm d}z(\rho) / {\rm d}\rho$ goes to infinity), or a conical lid (type [C]; the slope approaches unit value at the uppermost point), or a smooth dome (or inverted tear-drop, type [S], with zero slope).  The first two basic shapes come in two variants each:  In the ``early'' (e) or ``closed-orbit'' version, a focal line runs through the centerline of the ornament ($\rho = 0$) all the way from the bottom to the top. They also harbor the endpoint of a ``snake'' closed orbit.  In the ``late'' (l) or no-closed-orbit variant, the focal line goes only part way through the ornament, from the cusp at its bottom to a point within the ornament where it connects to an upward-pointing cusp from another ornament.  The three teardrop--shaped caustic subtypes share the latter property; they differ only in their parametrization. All seven types are shown in Fig.~\ref{fig:Caustic4.1}.

\subsubsection{Onion-domes}
\label{subsubsec:Caustic4.1}

\paragraph{Late variant [Ol].}
The caustic created in the $k$th cyclotron orbit $\bigl((k-1)\pi \leq t \leq k\pi\bigr)$ belongs to this class if $t > \frac\eta2$ holds for all points of the caustic.  In other words, the $k$th caustic surface is a late onion-dome if $\eta < 2(k-1)\pi$.  Its parametrization is simple:  The caustic starts at $t_0 = (k-1)\pi$ from its upper cusp at $\rho_0 = 0$, $z_0 = t_0(t_0 - \eta) / \eta$ (case 2e). As $t$ increases, $\rho_-(t)$ and $z_-(t)$ trace out the upper surface of the ``onion.''  When $t$ approaches $(k - \frac12)\pi$, $\rho_-(t)$ tends towards its maximum value $\rho =1$ (case 2a).  The ($-$) solution ceases to exist there, and the $(+)$ solution smoothly takes over.  It then traces out the lower part of the surface down to the lower cusp point $t_0 = k\pi$, $\rho_0 = 0$, $z_0 = \frac{k\pi}\eta(k\pi + \eta)$ (case 2e). From there, a focal line of length $2k\pi$ runs back along the symmetry axis to the upper cusp $z_0 = \frac{k\pi}\eta(k\pi - \eta)$ of the next late onion-dome, creating an infinite chain of overlapping surfaces linked by focal lines.

\paragraph{Early variant [Oe].}
If $t < \frac\eta2$ holds for all points of the $k$th caustic surface, i.e., $\eta > 2k\pi$, the caustic has the same shape, but a more complex temporal parametrization:  We begin at $t_0 = k\pi$ with the $(-)$ solution at the top cusp of the onion at $z_0 = \frac{k\pi}\eta(k\pi-\eta)$.  However, we now allow $t$ to \emph{decrease}.  When $t$ reaches $(k - \frac12)\pi$, we have arrived at the endpoint of a ``snake'' closed orbit that returns to the origin (case 2b).  To follow the curve smoothly, we switch to the $(+)$ solution, and then continue downward to $t_0 = t_{\rm min}^{(k)}$, the minimal time of flight to the caustic (case 2d).  There we switch back to the $(-)$ solution, and now allow $t$ to \emph{increase}.  When $t$ approaches $(k-\frac12)\pi$, we are at the maximum radial extension $\rho = 1$ (case 2a).  We switch again to the $(+)$ solution which describes the entire lower portion of the caustic down to the bottom cusp at $z_0 = \frac{k\pi}\eta(k\pi+\eta)$ that is attained for $t_0=k\pi$ (case 2e).  Note that the parametrization $\bigl[ \rho(t), z(t) \bigr]$ for this type is a \emph{double-valued} function of $t$.

Again, a focal line of length $2k\pi$ runs along the center of the ornament, but in this case it connects the opposite cusps of the caustic itself.  Early caustics, in contrast to their late counterparts, are isolated structures --- no continuous curve $(t, z, \rho)$ on the caustic set will connect them to another caustic surface.

\subsubsection{Conical lids}
\label{subsubsec:Caustic4.2}

Apart from the cone-like shape of the caustic at the upper end point (case 3a), they share all other aspects with the onion-dome caustics discussed above.  Conical lids occur only when $\eta$ is an integer multiple of $2\pi$.  For $\eta = 2k\pi$, the $(k+1)$th caustic surface turns into the late variant of the conical lid [Cl] that has the pattern of a late onion-dome, and is linked to their sequence via a focal line segment.  Simultaneously, the $k$th caustic becomes a conical lid in its early variant [Ce], a self-contained caustic with an associated closed orbit that is patterned after the early onion-domes.  The caustics are joined at the tip of the umbilic double cone structure.  A close-up is shown in Fig.~\ref{fig:Caustic3.1} (Appendix).

\subsubsection{Smooth domes (inverted tear-drops)}
\label{subsubsec:Caustic4.3}

For $2(k-1)\pi < \eta < 2k\pi$, the $k$th turning surface (with times of flight between $(k-1)\pi$ and $k\pi$) becomes a smooth-dome caustic that forms the uppermost boundary of classical motion, terminating on the symmetry axis for $t_0 = \frac\eta2$ at $z_0=-\frac\eta4$.  We may discern three types that interpolate between the early and late regimes of the motion (Fig.~\ref{fig:Caustic4.1}, bottom row).  In each case, a focal line segment joins the lower cusp of the caustic to the uppermost cusp of a late onion-dome surface.

\paragraph{Late variant [Sl].}
This case holds for $2(k-1)\pi < \eta \leq \eta_{\rm crit}^k$ (\ref{eq:Caustic3.0a}). We start with the $(-)$ solution from the top of the dome ($t_0=\frac\eta2$, case 2c).  As $t$ increases, we trace out the upper surface, and when $t$ goes through $(k-\frac12)\pi$, and $\rho$ reaches its maximum (case 2a), we smoothly switch to the $(+)$ solution that traces out the lower surface. The caustic curve is a single-valued function of $t$.

\paragraph{Intermediate variant [Si].}
Here, $\eta_{\rm crit}^k < \eta \leq (2k-1)\pi$.  We now start at $t_0=\frac\eta2$ from the uppermost point with the $(+)$ solution.  As $t$ decreases to $t_{\rm min}^{(k)}$ (case 2d) we trace out part of the smooth dome.  Then we switch to the $(-)$ solution and allow $t$ to increase again to $(k-\frac12)\pi$ (case 2a), completing the upper half of the caustic.  Switching back to the $(+)$ solution we trace the lower part of the ornament down to the cusp for $t_0=k\pi$.

\paragraph{Early variant [Se].}
In this case we again begin with the $(-)$ solution at the dome top ($t_0 = \frac\eta2$), but now $(2k-1)\pi < \eta < 2k\pi$ holds.  As $t$ decreases through $(k-\frac12)\pi$ (case 2b), we switch smoothly from the $(-)$ branch to the $(+)$ branch at the endpoint of a snake-shaped closed orbit.  The time $t$ continues to decrease, and construction of the remaining part of the caustic follows the pattern of the intermediate case sketched above.

\subsection{Evolution of the caustic pattern}
\label{subsec:Caustic5}

\begin{figure*}
\begin{center}
\includegraphics[width=\textwidth]{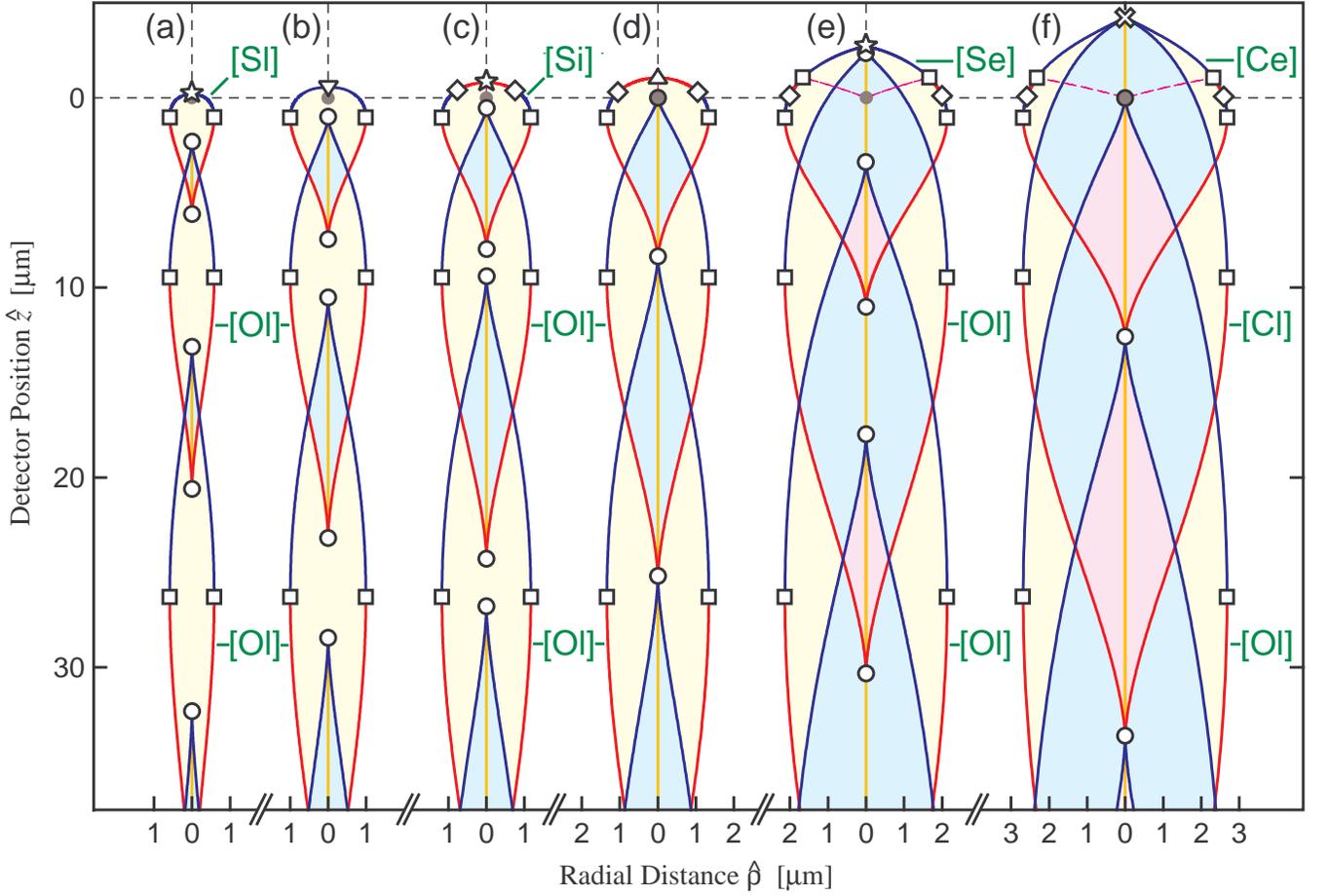}
\end{center}
\caption{Evolution of the caustic surface in parallel fields ${\cal E}= 15\,$V/m and ${\cal B} = 0.02\,$T.  Left to right: Emission energy $E = 3\,\mu$eV, $8.690\,\mu$eV, $12\,\mu$eV, $15.78\,\mu$eV, $40\,\mu$eV, and $63.13\,\mu$eV, corresponding to $\eta=1.370$, $\eta_{\rm crit}^{1} \approx 2.331$, $2.739$, $\pi$, $5.001$, and $2\pi$, respectively. --- Red and blue segments denote the ($+$) and ($-$) solution in the parametrization (\protect{\ref{eq:Caustic1.3}}), while focal lines connecting the caustics are shown in orange. Purple (dashed) curves:  Snake-type closed orbits.  Symbols indicate irregular points of the parametrization:  Half-integer orbits ($\Box$, cases 2a, 2b), top of caustic ($\boldsymbol\star$, case 2c), minimal time of flight ($\Diamond$, case 2d), cylindrical cusps ($\circ$, case 2e), umbilical points ($\boldsymbol\times$, case 3a), bifurcation point ($\bigtriangledown$, case 3b), and creation of a closed orbit ($\bigtriangleup$, case 3c).  The dots ($\bullet$) mark the source location. Yellow (blue, purple) shaded areas represent destination points connected to the source by $N=2$ (4, 6) classical trajectories.  Caustics are labeled according to their classification (green). --- (The caustic surface with $E=100\,\mu$eV ($\eta = 7.908$) depicted in Fig.~\protect{\ref{fig:Primer0.1}} would present the next step (g) in the sequence shown.)
\label{fig:Caustic5.1}
}
\end{figure*}

Considering the simplicity of the trajectories, the caustic surfaces in the parallel field problem are of surprising complexity.  Let us now examine how the caustic pattern evolves as we increase the energy of the emitted electron, or, more generally, increase the parameter $\eta = v_0{\cal B}/{\cal E}$.  Starting with $\eta \approx 0$, we obtain the sequence of changes displayed in Fig.~\ref{fig:Caustic5.1}.
\setcounter{paragraph}{0}
\begin{table}[t]
\begin{tabular}{c | c | c | c | c | c | c | c}
 & Ol & Cl & Sl & Si & Se & Ce & Oe \\ 
\hline
Closed orbit (snake)   & $-$     & $-$     & $-$     & $-$     & $\surd$ & $\surd$ & $\surd$ \\
Closed orbit (balloon) & $-$     & $-$     & $-$     & $-$     & $-$     & $-$     & $\surd$ \\
isolated caustic       & $-$     & $-^*$   & $-$     & $-$     & $-$     & $\surd^*$&$\surd$ \\
downhill motion        & $\surd$ & $\surd$ & $\surd$ & $\surd$ & $-$     & $-$     & $-$     \\
upper turning point    & $-$     & $\surd$ & $\surd$ & $\surd$ & $\surd$ & $\surd$ & $-$     \\
off-axis minimum in $t$& $-$     & $-$     & $-$     & $\surd$ & $\surd$ & $\surd$ & $\surd$ \\
\hline
succeeding type & Cl & Sl & Si & Se & Ce & Oe & Oe \\
\hline
\end{tabular}
\caption{Summary of properties of the caustics listed in Fig.~\ref{fig:Caustic4.1}.  The ``early'' or ``closed-orbit'' types [Se], [Ce], and [Oe] contain the endpoint of a snake-type closed orbit.  Early onion-domes [Oe] are additionally associated with a balloon-type closed orbit.  This type is also isolated; the focal line associated with it joins its two cusp singularities on the symmetry axis instead of connecting the lower cusp to the opposite cusp of the following ``late'' onion-dome [Ol].  (The conical lids [Cl], [Ce] marked by $*$ are linked at the umbilical point.)  The late types [Ol], [Cl], [Sl], together with the intermediate case [Si], all share downhill motion, i.e., $\dot z(t) \geq 0$ (\protect{\ref{eq:Class2.8}}) holds for all points of the caustic.  The uphill turning point belongs either to one of the smooth dome caustics [Sl]--[Se] or to the pair of conical lid types [Cl], [Ce].  For the types [Si] through [Oe], the minimum time of flight to the caustic occurs off the symmetry axis, while the late types [Ol]--[Sl] are single-valued functions of $t$. --- The last row illustrates that the caustic types in the given order form an evolutionary sequence:  With increasing $\eta$, each caustic is transformed in six successive steps from a late onion-dome [Ol] to an early onion-dome [Oe].
\label{tab:Caustic5.1}}
\end{table}

\paragraph{$0 < \eta < \eta_{\rm crit}^{1}$.}
($\eta_{\rm crit}^{1} \approx 2.331$ is defined in Eq.~(\ref{eq:Caustic3.0a}).)  The top of the first caustic ($k=1$) is a smooth dome, and the curve is parametrized in the way of the late variant [Sl].   Its lowest point is a downhill-pointing cusp corresponding to $t_0=\pi$ at $z_0 = \frac\pi\eta(\pi + \eta)$ (Eq.~(\ref{eq:Class1.3})).  Extending upward from this cusp is a focal line, which ends at $z_0 = \frac\pi\eta(\pi-\eta)$.  This is the uppermost point of an onion-dome caustic that begins here in an uphill-pointing cusp.  Its parametrization is again the late variant [Ol].  Below it hang an infinite sequence of similar onion-domes interconnected by focal lines running from the lowest point of each caustic at $z_0 = \frac{k\pi}\eta(k\pi + \eta)$ to the highest point of the next one at $z_0 = \frac{k\pi}\eta(k\pi - \eta)$.  Near these focal lines, four trajectories go through every point on the detector.

\paragraph{$\eta = \eta_{\rm crit}^{1}$.}
We are at the first bifurcation point (case 3b).  The shape stays the same, but the parametrization of the smooth dome changes here.

\paragraph{$\eta_{\rm crit}^{1} < \eta < \pi$.}
The parametrization of the smooth dome is now the intermediate-time variant [Si], with a newly created off-axis minimum (case 2d) in the time of flight.  As $\eta$ increases, the lengths of the onion-dome caustics increase, with their upper cusps moving up and their lower cusps moving down.

\paragraph{$\eta = \pi$.}
Several things happen here.  The top of the highest onion-dome moves up to the origin, and a snake-type closed orbit is created (case 3c).  The bottom cusp of each ornament touches the cusp at the top of the second ornament down.  Therefore, focal lines now cover the entire positive $z$ axis.  The parametrization of the smooth dome changes again.

\paragraph{$\pi < \eta < 2\pi$.}
The uppermost caustic is now parametrized by the early variant [Se] of the smooth dome.  The first snake-type closed orbit is present, and the cusp that had touched the source has moved further uphill.  New diamond-shaped regions appear in which six trajectories lead to every point.  In these regions, the focal line at $\rho=0$ may be said to be degenerate:  There are two distinct families of trajectories that focus at the symmetry axis.

\paragraph{$\eta = 2\pi$.}
We have reached the umbilical point (case 3a).  The uppermost cusp now has caught up with the uppermost point of the smooth dome.  The caustics involved in the ``collision'' have been deformed into conical lids --- the former early-type smooth dome has turned into its early variant [Ce], while the late-type onion dome takes on the late-type parametrization [Cl].  Another cusp has moved up to touch the origin, and bottom and top cusps of different ornaments again coincide.  This is the bifurcation point for a balloon-type closed orbit.

\paragraph{$2\pi < \eta < \eta_{\rm crit}^{2}$.}
This situation is depicted in Fig.~\ref{fig:Primer0.1}.  The former onion dome ($k=2$) has moved to the top of the caustic structure and has further evolved into a smooth-dome with late-type parametrization [Sl], while the original smooth dome ($k=1$) has become the first onion-dome belonging to the early variant [Oe].  The ballon-type closed orbit has detached from the symmetry axis, and diamond-shaped regions with eightfold path degeneracy occur.  Apart from the presence of the innermost [Oe] caustic, the caustic pattern is the same as in step (a).\\

At this point, the evolutionary cycle (with a period $2\pi$ in $\eta$) starts over:  As $\eta$ grows further, the pattern continues to evolve in the same manner.  Each time $\eta$ increases through $\pi$, a new closed orbit is generated, alternating snakes ($\eta = (2k-1)\pi$) and balloons ($\eta = 2k\pi$), and the maximum number of classical paths leading from the origin to any off-axis destination ($\rho \neq 0$) increases by two.  The $k$th caustic surface undergoes its evolution in the interval $2(k-1)\pi \leq \eta \leq 2k\pi$ that ultimately results in the transformation of an onion dome [Ol] with the late parametrization scheme into an onion dome of the early variant [Oe].  --- The properties of the different types and variants of caustics in parallel fields, and their evolution with increasing $\eta$, are briefly summarized in Table~\ref{tab:Caustic5.1}.

\section{Semiclassical and uniform approximations}
\label{sec:Semi}

Except at the caustic surfaces, the quantum wave function $\psi(\mathbf{\hat r})$ generated by a fixed energy point source in parallel fields is well approximated by a sum over all $N(\mathbf{\hat r})$ classical paths connecting the source with the destination point $\mathbf{\hat r}$ on the detector \cite{Berry1972a}:
\begin{equation}
\label{eq:Semi0.1}
\psi_{\mathrm{sc}}(\mathbf{\hat r}) = \sum_{\alpha=1}^{N} \psi_\alpha(\mathbf{\hat r}) \;.
\end{equation}
The magnitude and phase of the semiclassical wave function for trajectory $\alpha$ depend on its classical density $\rho_\alpha = \rho_{\rm cl}^{(\alpha)}(\mathbf{\hat r})$ and classical action functional ${\cal W}_\alpha(\mathbf{\hat r};E)$ (\ref{eq:Class2.13}), respectively \footnote{To achieve agreement with the Green function we introduce a (physically irrelevant) overall phase factor $(-1)$.}:
\begin{equation}
\label{eq:Semi0.2}
\psi_\alpha(\mathbf{\hat r}) = - \sqrt{\rho_\alpha(\mathbf{\hat r})} \, \exp\bigl({\rm i}\chi_\alpha(\mathbf{\hat r}) \bigr) \;.
\end{equation}
Here, the phase $\chi_\alpha(\mathbf{\hat r})$ is given by:
\begin{equation}
\label{eq:Semi0.3}
\chi_\alpha(\mathbf{\hat r}) = \frac1\hbar {\cal W}_\alpha(\mathbf{\hat r};E) - \frac\pi2 \mu_\alpha(\mathbf{\hat r}) \;.
\end{equation}
The \emph{Maslov index} $\mu_\alpha(\mathbf{\hat r})$ keeps track of the singularity ``history'' of the trajectory \cite{Maslov1981a}.  A more accurate result is obtained if we extend the sum (\ref{eq:Semi0.1}) to include \emph{tunneling trajectories}.  Their times of flight $T(\mathbf r;E)$ are complex roots of Eq.~(\ref{eq:Class2.12}); their action ${\cal W}_\alpha(\mathbf{\hat r};E)$ (\ref{eq:Class2.13}), formally obtained by analytic continuation, has an imaginary part, so their contribution to the semiclassical wave function is exponentially suppressed.

The wave function (\ref{eq:Semi0.1}) leads to semiclassical expressions for the particle density $\rho_{\rm sc}(\mathbf{\hat r})$ and current density distribution $\mathbf j_{\rm sc}(\mathbf{\hat r})$.  Ignoring tunneling trajectories, they differ from their classical counterparts by the presence of interference terms \footnote{We also omit contributions of order ${\cal O}(\hbar)$.}:
\begin{gather}
\label{eq:Semi0.4a}
\rho_{\rm sc}(\mathbf{\hat r}) = \sum_{\alpha=1}^{N} \rho_\alpha + 2\sum_{\alpha < \beta} \sqrt{\rho_\alpha \rho_\beta} \,\cos(\chi_\alpha - \chi_\beta) \;, \\
\label{eq:Semi0.4b}
\mathbf j_{\rm sc}(\mathbf{\hat r}) = \sum_{\alpha=1}^{N} \mathbf j_\alpha + \sum_{\alpha < \beta} \sqrt{\rho_\alpha \rho_\beta} \, (\mathbf v_\alpha + \mathbf v_\beta) \cos( \chi_\alpha - \chi_\beta) \;.
\end{gather}
Here, $\mathbf v_\alpha$ is the electron velocity at the detector $\mathbf{\hat r}$, and $\mathbf j_\alpha = \rho_\alpha \mathbf v_\alpha$ denotes the local classical current density along trajectory $\alpha$. 

\subsection{Classical density and current}
\label{subsec:Semi1}

To find the WKB wave function, we now determine the particle and current densities $\rho_\alpha(\mathbf{\hat r})$, $\mathbf j_\alpha(\mathbf{\hat r})$ for the classical trajectories $\alpha$ in parallel fields.  A bundle of trajectories emitted from the source under spherical angles $\theta'$ and $\phi'$ (\ref{eq:Class2.6}) will cross through the detector surface (here a perpendicular section with fixed $\hat z$) at a radial distance $\hat\rho = \rho d$ (\ref{eq:Class0.2}) and a polar angle $\phi$ (\ref{eq:Class1.4}).  Particle conservation then requires that the emission rate at the source:
\begin{equation}
\label{eq:Semi1.1}
J_\alpha = J_0 f(\theta'_\alpha,\phi'_\alpha) \sin\theta'_\alpha {\rm d}\theta'_\alpha {\rm d}{\phi'_\alpha} \;,
\end{equation}
equals the flux through the detector surface (with normal vector $\mathbf a$):
\begin{equation}
\label{eq:Semi1.2}
J_\alpha = \mathbf j_\alpha(\mathbf{\hat r}) \cdot{\rm d}\mathbf a 
= j_z^{\alpha}(\mathbf{\hat r}) \hat\rho{\rm d}\hat\rho{\rm d}\phi 
= \frac{v_0^2}{2\omega_L^2} j_z^{\alpha}(\mathbf{\hat r}){\rm d}(\rho^2){\rm d}\phi \;.
\end{equation}
$J_0$ denotes the total emission rate which we set equal to the total current generated by an isotropic quantum point source $C \delta^3(\mathbf{\hat r})$ of free electrons with wave number $k=(2mE)^{1/2}/\hbar$ \cite{Wigner1948a}:
\begin{equation}
\label{eq:Semi1.2a}
J_0(E) = |C|^2 \frac{mk}{\pi\hbar^3} \;.
\end{equation}
(See Section~\ref{subsec:Quantum3}.)  $f(\theta',\phi')$ is the (normalized) angular distribution of the source.  We assume isotropic emission, i.e., an s--wave source with $f(\theta',\phi') = 1/(4\pi)$.

Equating (\ref{eq:Semi1.1}), (\ref{eq:Semi1.2}) we find for the current density $j_z^\alpha$:
\begin{equation}
\label{eq:Semi1.3}
j_z^{\alpha}(\mathbf{\hat r}) = \frac{2\omega_L^2}{v_0^2} J_0 f(\theta'_\alpha,\phi'_\alpha) \left| \frac{\partial(\cos\theta'_\alpha,\phi'_\alpha)}{\partial(\rho^2,\phi)} \right|
\;.
\end{equation}
The relationship between the emission angle and the location on the detector plane is given by Eq.~(\ref{eq:Class2.6}), where we have to keep in mind that the time of flight $T = T(\mathbf r;E)$ in this formula is a function of the distances $\rho$ and $z$, and the energy $E$.  Because $\partial\phi'/\partial\phi = 1$ and $\partial\cos\theta'/\partial\phi = 0$ the Jacobian in Eq.~(\ref{eq:Semi1.3}) reduces to a simple derivative:
\begin{equation}
\label{eq:Semi1.4}
j_z^{\alpha}(\mathbf{\hat r}) = \frac{2\omega_L^2}{v_0^2} J_0 f(\theta'_\alpha,\phi'_\alpha) \left| \frac{\partial\cos\theta'_\alpha}{\partial\rho^2} \right| \;.
\end{equation}
To evaluate it, we note that (from Eqs.~(\ref{eq:Class2.3}) and (\ref{eq:Class2.6}))
\begin{equation}
\label{eq:Semi1.5}
\frac{\partial\cos\theta'}{\partial\rho^2} = - \Bigl( \frac z{T^2} + \frac1\eta \Bigr) \frac{\partial T}{\partial\rho^2} = - \frac{\dot z}T \frac{\partial T}{\partial\rho^2} \;.
\end{equation}
Implicit differentiation of $\epsilon(T)$ (\ref{eq:Class2.12}) with respect to $\rho^2$ yields:
\begin{equation}
\label{eq:Semi1.6}
\frac1{\sin^2 T} + \frac{\partial\epsilon}{\partial T}\,\frac{\partial T}{\partial\rho^2} = 0 \;.
\end{equation}
Finally, the particle velocity in the $z$ direction is given by ${\rm d}\hat z/{\rm d}\hat t = v_0 \dot z$ (\ref{eq:Class2.8}).  Combining (\ref{eq:Semi1.4})--(\ref{eq:Semi1.6}) we find for the classical density:
\begin{equation}
\label{eq:Semi1.7}
\rho_\alpha(\mathbf{\hat r}) = \frac{j_z^\alpha}{v_0\dot z_\alpha} = \frac{2\omega_L^2 J_0}{v_0^3} \frac{f(\theta'_\alpha,\phi'_\alpha)}{T_\alpha \sin^2 T_\alpha \bigl| \frac{\rho^2\cos T_\alpha}{\sin^3 T_\alpha} + \frac{z^2}{T_\alpha^3} - \frac {T_\alpha}{\eta^2} \bigr|} \;.
\end{equation}
This expression diverges at the caustics where $\partial\epsilon/\partial T$ (\ref{eq:Caustic1.2}) vanishes.  It is also singular on the focal line segments linking cusps of caustic surfaces (Section~\ref{subsec:Caustic3}) as $T=k\pi$ holds there.

\subsection{The Maslov index}
\label{subsec:Semi2}

It remains to determine the \emph{Maslov index} $\mu_\alpha(\mathbf{\hat r})$ of the trajectories.  This integer tracks the sign changes of the determinant in Eq.~(\ref{eq:Semi1.3}) along the classical path \cite{Schulman1981a,Maslov1981a,Delos1986a} --- in other words, the number of points along the trajectory $\alpha$ where the density $\rho_\alpha$ (\ref{eq:Semi1.7}) diverges.  In the parallel field problem, this happens (i) when the trajectory traverses a focal line segment ($t=j\pi$, where $0 < t < T$ is the time of flight along the path); and (ii) when the trajectory is reflected at a caustic surface, i.e., the function $\frac{\partial\epsilon}{\partial t}$ becomes zero along the path.  Inserting the equation of motion (\ref{eq:Class1.3}), (\ref{eq:Class1.4}) into Eq.~(\ref{eq:Caustic1.2}) we obtain the equivalent condition:
\begin{equation}
\label{eq:Semi2.1}
\sin^2\theta'\cot t + \frac2\eta \cos\theta' + \frac1t \cos^2\theta' = 0 \;.
\end{equation}
Within each cyclotron period $(k-1)\pi < t < k\pi$, the left-hand side of this expression is a strictly monotonic function that approaches $\pm\infty$ at the interval boundaries:  For a given emission angle $\theta'$ there is a single solution to Eq.~(\ref{eq:Semi2.1}) in each period, so \emph{every trajectory eventually touches every caustic surface exactly once}.  (Note that the implicit equation (\ref{eq:Semi2.1}) is equivalent to the parametrization of $\cos\theta_\pm'(t)$ (\ref{eq:Caustic1.4}) obtained earlier.)

Figure~\ref{fig:Class3.1} illustrates that for the pair of trajectories with times of flight $(k-1)\pi < t < k\pi$, $\frac{\partial\epsilon}{\partial t}$ is negative for the early solution $T^{(k)}_<$ but positive for the late solution $T_>^{(k)}$ of Eq.~(\ref{eq:Class2.12}).  The change of sign implies that the late path, unlike the early path, has been reflected off the $k$th caustic surface.  Since both trajectories have crossed $(k-1)$ focal lines and turned at $(k-1)$ caustics before, we find for their respective Maslov index:
\begin{equation}
\label{eq:Semi2.2}
\mu_\alpha(\mathbf{\hat r}) = 
\begin{cases}
2k-2 & \qquad T_\alpha = T_<^{(k)} \;,\\
2k-1 & \qquad T_\alpha = T_>^{(k)} \;.
\end{cases}
\end{equation}

\subsection{Caustics and a uniform approximation}
\label{subsec:Semi3}

The semiclassical approximation (\ref{eq:Semi1.7}) generally performs remarkably well, but it diverges for those trajectories that cross over each other at a caustic.  To correct the divergent behavior, we may replace the semiclassical contribution of the offending classical paths in the sum (\ref{eq:Semi0.4a}) by \emph{uniform approximations}, smooth functions that describe the refracting electron wave near the caustic.  \emph{Catastrophe Theory} \cite{Thom1975a,Berry1976a,Poston1978a} offers a mathematical formalism to classify the points of the caustic set in terms of a small group of generic cases, the ``catastrophes.'' 
\begin{table}[t]
\begin{tabular}{c | c}
Type & Generic Catastrophe\\ 
\hline
Caustic Surface & fold \\
Focal Line & --- \\
Cylindrical Cusp & cusp \\
Conical Lid & hyperbolic umbilic \\
Double Cusp & hyperbolic umbilic \\
\end{tabular}
\caption{Catastrophe classification of caustic points in parallel fields.  All points of the caustic suface with $\rho > 0$ belong to the simplest class, the fold; a sequence of higher catastrophes is found on the symmetry axis.  (Double cusp catastrophes occur only in the limit $E\rightarrow 0$ and are not discussed in this paper.)
\label{tab:Semi3.1}}
\end{table}
Several examples of higher-order catastrophes occur in the parallel field problem, and we list them in Table~\ref{tab:Semi3.1}.  They are located on the symmetry axis ($\rho = 0$).

In contrast, the caustic surfaces off the symmetry axis ($\rho > 0$) all represent the most basic catastrophe, the \emph{fold}.  Electron waves undergoing reflection at the fold-type caustic surfaces locally resemble Airy functions \cite{Connor1976a,Poston1978a,Abramowitz1965a}.  This suggests an approximation scheme that works uniformly for all $\rho > 0$:  In each cyclotron period $(k-1)\pi < T < k\pi$ we replace the semiclassical contributions $\psi_\alpha(\mathbf r) = \psi_{<,>}^{(k)}(\mathbf r)$ (\ref{eq:Semi0.2}) due to the two trajectories with time of flight $T^{(k)}_<$, $T^{(k)}_>$ in this interval (see Fig.~\ref{fig:Class3.1}) by a linear combination of Airy functions:
\begin{equation}
\label{eq:Semi3.1}
\psi_{\rm uni}^{(k)}(\mathbf r) = 
\gamma_k(\mathbf r) \Ai\bigl(u_k(\mathbf r)\bigr) + 
\delta_k(\mathbf r) \Ai'\bigl(u_k(\mathbf r)\bigr) \;.
\end{equation}
To obtain the uniform approximation we then sum $\psi_{\rm uni}^{(k)}(\mathbf r)$ over all cyclotron orbits $k$.

We choose the auxiliary functions in Eq.~(\ref{eq:Semi3.1}) so that after asymptotic expansion of the Airy functions \cite{Abramowitz1965a}, the leading term matches the combined semiclassical contributions of the two classical paths in the $k$th cyclotron period:  The argument of the Airy functions $u_k(\mathbf r)$ only depends on the difference in the classical action of the two paths, $\Delta {\cal W}_k = {\cal W}_{k,>} - {\cal W}_{k,<}$:
\begin{equation}
\label{eq:Semi3.2}
u_k(\mathbf r) = - \left( \frac{3\Delta{\cal W}_k(\mathbf r;E)}{4\hbar} \right)^{2/3} \;,
\end{equation}
while the ``amplitudes'' $\gamma_k(\mathbf r)$ and $\delta_k(\mathbf r)$ are functions of the classical densities $\rho_\alpha(\mathbf r)$ (\ref{eq:Semi1.7}) along these trajectories:
\begin{align}
\label{eq:Semi3.3a}
\gamma_k &= (-1)^k \sqrt\pi |u_k|^{1/4} 
{\rm e}^{{\rm i}\overline{\cal W}_k/\hbar - {\rm i}\pi/4}
\bigl( \sqrt{\rho_{k,>}} + \sqrt{\rho_{k,<}} \bigr)\;, \\
\label{eq:Semi3.3b}
\delta_k &= (-1)^k \sqrt\pi |u_k|^{-1/4} 
{\rm e}^{{\rm i}\overline{\cal W}_k/\hbar + {\rm i}\pi/4}
\bigl( \sqrt{\rho_{k,>}} - \sqrt{\rho_{k,<}} \bigr) \;.
\end{align}
Here, $\overline{\cal W}_k = \frac12 \bigl({\cal W}_{k,>} + {\cal W}_{k,<}\bigr)$ denotes the average action for both paths.  A similar identification scheme applies in the case of classically forbidden motion, where we use a tunneling trajectory to fix $u_k(\mathbf r)$, $\gamma_k(\mathbf r)$, and $\delta_k(\mathbf r)$ instead.  These auxiliary functions, and therefore the uniform approximation (\ref{eq:Semi3.1}), behave smoothly at the caustic.  

As noted before, the approximation is only valid for fold catastrophes, and, like the semiclassical solution, fails near the focal lines, cusps, and umbilic catastrophes that occur on the symmetry axis.  These regions require a separate treatment.

\section{The Quantum Solution}
\label{sec:Quantum}

In the quantum description, the motion of the electrons in the external potential $\mathbf A(\mathbf{\hat r})$, $\Phi(\mathbf{\hat r})$ (\ref{eq:Class0.1}) is controlled by the Hamiltonian ${\cal H}$ \footnote{For simplicity, we ignore the electron magnetic moment $\mu$ that merely leads to an effective shift $\pm\frac12 g\mu_B {\cal B}$ in the electron energy, depending on its spin orientation.}:
\begin{equation}
\label{eq:Quantum0.1}
{\cal H}(\mathbf{\hat r}, \mathbf{\hat p}) = 
\frac{{\rm d}\mathbf{\hat r}}{{\rm d}\hat t} \cdot \mathbf{\hat p} - 
{\cal L}(\mathbf{\hat r}, \mathbf{\hat p}) =
\frac 1{2m} \bigl( \mathbf{\hat p} - q\mathbf A(\mathbf{\hat r}) \bigr)^2 + q\Phi(\mathbf{\hat r}) \;.
\end{equation}
As in Section~\ref{sec:Classical}, we again introduce dimensionless variables.  A natural energy scale $\epsilon$ for the perpendicular motion is given by:
\begin{equation}
\label{eq:Quantum0.2}
\epsilon = E/(\hbar\omega_L) \;,
\end{equation}
and $\cal H$ takes on a particularly simple form if we set:
\begin{equation}
\label{eq:Quantum0.3}
\xi = \sqrt{\frac{m\omega_L}\hbar} \hat x = \sqrt{2\epsilon} x \,, \quad
\upsilon = \sqrt{\frac{m\omega_L}\hbar} \hat y = \sqrt{2\epsilon} y \;,
\end{equation}
For the motion in the field direction, we define an inverse energy scale $\beta$:
\begin{equation}
\label{eq:Quantum0.4}
\beta = \bigl( \frac{2m}{\hbar^2 q^2 {\cal E}^2} \bigr)^{1/3} \;,
\end{equation}
and a dimensionless coordinate $\zeta$:
\begin{equation}
\label{eq:Quantum0.5}
\zeta = \beta q{\cal E} \hat z = 2 \bigl( \frac{2\epsilon^2}\eta \bigr)^{1/3} z \;.
\end{equation}

Because the motion along the symmetry axis and in the perpendicular $x$--$y$ plane are independent, the Hamiltonian ${\cal H}$ (\ref{eq:Quantum0.1}) is separable:
\begin{equation}
\label{eq:Quantum1.1}
{\cal H}(\mathbf{\hat r}, \mathbf{\hat p}) = 
{\cal H}_\perp(\mathbf {\hat r}_\perp, \mathbf {\hat p}_\perp) + 
{\cal H}_\parallel(\hat z, \hat p_z) \;,
\end{equation}
and the partial Hamiltonians commute, $[{\cal H}_\perp,{\cal H}_\parallel] = 0$.  Using dimensionless variables, they read:
\begin{multline}
\label{eq:Quantum1.2}
{\cal H}_\perp(\mathbf {\hat r}_\perp, \mathbf {\hat p}_\perp) = \hbar\omega_L \Bigl[ 
-\frac12 \bigl( \partial_\xi^2 + \partial_\upsilon^2 \bigr) \Bigr. \qquad \\
\qquad \Bigl. + {\rmi} \bigl(\xi\partial_\upsilon - \upsilon\partial_\xi \bigr) + \frac12 \bigl( \xi^2 + \upsilon^2 \bigr) \Bigr] \,,
\end{multline}
and 
\begin{equation}
\label{eq:Quantum1.3}
{\cal H}_\parallel(\hat z, \hat p_z) = -\frac1\beta \bigl[ \partial_\zeta^2 + \zeta \bigr] \;.
\end{equation}

\subsection{Electronic eigenfunctions}
\label{subsec:Quantum1}

\begin{figure}
\begin{center}
\includegraphics[width=\columnwidth]{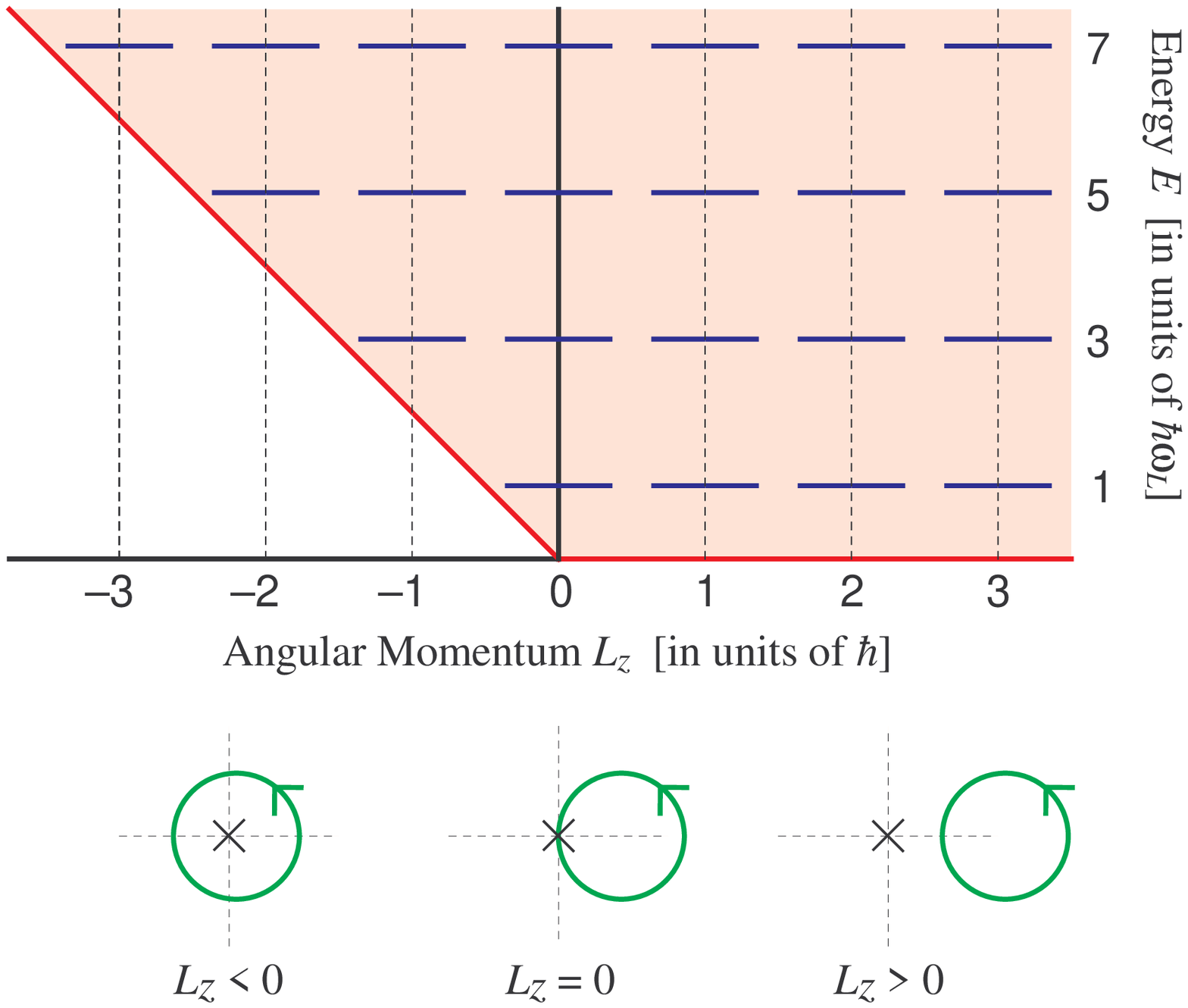}
\end{center}
\caption{Top: The spectrum of ${\cal H}_\perp$.  The eigenstates can be ordered by their angular momentum $L_z = \mu\hbar$, and their energy $E_\nu = (2\nu+1)\hbar\omega_L$.  Their distribution discretizes the classically allowed range $E \geq 0$, $E \geq -2\omega_L L_z$ (red area). --- Bottom: Corresponding classical orbits in the $x$--$y$ plane. The energy $E = 2m\omega_L^2 R^2$ determines their radius $R$, while the angular momentum $L_z$ contains information about their position relative to the origin:  The orbit encircles the origin for $L_z < 0$, while the origin is outside the orbit for $L_z > 0$.  $L_z$ vanishes for orbits passing through the origin (like the trajectories of Section~\protect{\ref{sec:Classical}}).  (The magnetic field points toward the reader.)
\label{fig:Quantum1.1}
}
\end{figure}

The Hamiltonian ${\cal H}_\perp$ (\ref{eq:Quantum1.2}) governs the cyclotron motion of an electron confined to the $x$--$y$ plane in a uniform magnetic field.  It differs from the Hamiltonian of a two-dimensional harmonic oscillator only by the middle term $-\omega_L {\rm L}_z$ which is constant for radially symmetric eigenstates with fixed angular momentum quantum number $\mu$.  The stationary states are thus oscillator eigenstates that are shifted in energy; we express them in terms of Laguerre polynomials $L_n^\mu(\bar\rho^2)$ \cite{Abramowitz1965a}, where $\bar\rho^2 = \xi^2 + \upsilon^2$ is the scaled radial distance \cite{Johnson1983a,Greene1987a}:
\begin{equation}
\label{eq:Quantum1.4}
\chi^\perp_{n,\mu}(\bar\rho,\phi) = \sqrt{\frac{m\omega_L}{\pi\hbar} \frac{n!}{(n+|\mu|)!}} {\bar\rho}^{|\mu|} L_n^{|\mu|}(\bar\rho^2) {\rm e}^{-\bar\rho^2/2} {\rm e}^{{\rm i}\mu\phi} \;.
\end{equation}
For $\mu \geq 0$, but not for negative $\mu$, their eigenenergy depends only on the principal quantum number $n=0,1,2,\ldots$:
\begin{equation}
\label{eq:Quantum1.5}
E^{\perp}_{n,\mu} = 
\begin{cases}
(2n+1)\hbar\omega_L  & (\mu \geq 0)\;, \\
(2n+2|\mu|+1)\hbar\omega_L  & (\mu < 0)\;.
\end{cases}
\end{equation}
The spectrum is discrete, and the states occupy equally spaced, infinitely degenerate \emph{Landau levels} $E_\nu = (2\nu + 1)\hbar\omega_L$ \cite{Landau1977a}.  The unusual skewed spectrum can be explained in terms of classical cyclotron orbits (see Fig.~\ref{fig:Quantum1.1}).

The Schr\"odinger equation for motion along the $z$ axis, $\bigl[E_\parallel - {\cal H}_\parallel\bigr]\chi_\parallel(\zeta;E_\parallel) = 0$, is essentially Airy's differential equation \cite{Landau1977a}, and the eigenfunctions can be expressed as regular Airy functions \cite{Abramowitz1965a}: 
\begin{equation}
\label{eq:Quantum1.8}
\chi^\parallel_{\rm reg}(\zeta;E_\parallel) = \Ai\bigl[-(\beta E_\parallel + \zeta) \bigr] \;.
\end{equation}
For our purposes, we will also need a Hankel--type solution of the differential equation that behaves like an outgoing wave in the asymptotic limit $z \rightarrow \infty$ \cite{Landau1977a,Bracher1998a,Kramer2002a}:
\begin{equation}
\label{eq:Quantum1.9}
\chi^\parallel_{\rm out}(\zeta;E_\parallel) = \Ci\bigl[-(\beta E_\parallel + \zeta) \bigr] \;.
\end{equation}
Here, $\Ci(u) = \Bi(u) + {\rm i}\Ai(u)$ is a linear superposition of the standard Airy functions \cite{Abramowitz1965a}.

\subsection{Construction of the Green function}
\label{subsec:Quantum2}

The wave function $\psi(\mathbf r)$ generated by a stationary, pointlike, isotropic source of electrons with energy $E$ at position $\mathbf{\hat r}'$ is a solution of the \emph{inhomogeneous} Schr\"odinger equation \cite{Kramer2002a}:
\begin{equation}
\label{eq:Quantum2.1}
\bigl[ E - {\cal H} \bigr]\psi(\mathbf{\hat r}, \mathbf{\hat r}'; E) = 
C \delta^{(3)}(\mathbf{\hat r} - \mathbf{\hat r}') \;.
\end{equation}
The parameter $C$ describes the strength of the electron source.  The solution $\psi$ of Eq.~(\ref{eq:Quantum2.1}) is then a multiple of the \emph{energy Green function} $G(\mathbf{\hat r},\mathbf{\hat r}';E)$ which is formally the resolvent operator (i.e., the inverse of $E - {\cal H}$) in position space \cite{Economou1983a}:
\begin{equation}
\label{eq:Quantum2.2}
G(\mathbf{\hat r},\mathbf{\hat r}';E) = \lim_{\tilde\eta \rightarrow 0^+}
\left\langle \mathbf{\hat r} \left| \bigl[E - {\cal H} + {\rm i}\tilde\eta\bigr]^{-1} \right| \mathbf{\hat r}' \right\rangle \;.
\end{equation}
The infinitesimal parameter $\tilde\eta$ ensures that the Green function asymptotically shows outgoing-wave behavior.

The quantum mechanical solution therefore requires knowledge of the Green function.  In three-dimensional space, explicit expressions for $G(\mathbf{\hat r},\mathbf{\hat r}';E)$ are available only for a handful of simple potentials \cite{Dalidchik1976a,Bracher1998a,Hostler1964a,Bakhrakh1971a}, not including the parallel fields Hamiltonian ${\cal H}$ (\ref{eq:Quantum1.1}). However, $G(\mathbf{\hat r},\mathbf{\hat r}';E)$ can be written as a series of readily available one-dimensional Green functions \cite{Bakhrakh1971a}:  We first expand Eq.~(\ref{eq:Quantum2.2}) into the complete set of eigenstates $\chi^\perp_{n,\mu}$ (\ref{eq:Quantum1.4}), (\ref{eq:Quantum1.5}) of the ``magnetic'' Hamiltonian ${\cal H}_\perp = {\cal H} - {\cal H}_\parallel$:
\begin{multline}
\label{eq:Quantum2.3}
G(\mathbf{\hat r},\mathbf{\hat r}';E) = 
\sum_{n,\mu} \chi^\perp_{n,\mu}(\bar\rho,\phi) \chi^\perp_{n,\mu}(\bar\rho',\phi')^* \\
\times\, \lim_{\tilde\eta \rightarrow 0^+} \left\langle \hat z \left| \bigl[E - E^\perp_{n,\mu} - {\cal H}_\parallel + {\rm i}\tilde\eta\bigr]^{-1} \right| \hat z' \right\rangle \;.
\end{multline}
Comparison with Eq.~(\ref{eq:Quantum2.2}) shows that the second line is actually the outgoing-wave Green function for an electron with effective energy $E^\parallel_{n,\mu} = E - E^\perp_{n,\mu}$ moving in the one-dimensional linear potential ${\cal H}_\parallel$ (\ref{eq:Quantum1.3}).  To obtain it in explicit form we match the regular and outgoing-wave solutions for a uniformly accelerated electron (\ref{eq:Quantum1.8}), (\ref{eq:Quantum1.9}) at the source $\hat z = \hat z'$ \cite{Bracher1998a}:
\begin{equation}
\label{eq:Quantum2.4}
G_\parallel(\zeta, \zeta'; E_\parallel) = -\pi\beta^2 q{\cal E} \Ai(-u_<) \Ci(-u_>) \;.
\end{equation}
Here, $u_< = \beta E_\parallel + \min(\zeta, \zeta')$ and $u_> = \beta E_\parallel + \max(\zeta, \zeta')$.

We now assume that the source is located at the origin, $\mathbf{\hat r}' = \mathbf 0$.
In this case, only magnetic eigenstates with $\mu=0$ contribute to the series (\ref{eq:Quantum2.3}) (all other eigenfunctions $\chi^\perp_{n,\mu}(\bar\rho',\phi')$  vanish for $\bar\rho'=0$).  Because $L_n^0(0) = 1$ \cite{Abramowitz1965a}, we find the simplified expansion \cite{Fabrikant1991a,Kramer2001a,Kramer2004a}:
\begin{multline}
\label{eq:Quantum2.5}
G(\mathbf{\hat r};E) = - \frac{m{\cal B}}{\beta\hbar^3 {\cal E}} {\rm e}^{-\bar\rho^2/2} 
\sum_{n=0}^\infty L_n(\bar\rho^2) \\ \times\, 
\Ai\bigl[-\beta (E - E_n) \bigr] \Ci\bigl[-\beta (E - E_n) - \zeta \bigr] \;,
\end{multline}
where $E_n = (2n + 1)\hbar\omega_L$ is the energy bound in the cyclotron motion (\ref{eq:Quantum1.5}).  The result holds for $\hat z \geq 0$; for $\hat z < 0$, the arguments of the Airy functions have to be exchanged.  

For large $n$, the energy $E_\parallel = E - E_n$ available for motion along the $z$ axis is negative, the electrons are forced to tunnel from $\hat z'=0$ to $\hat z$, and the corresponding term in the expansion (\ref{eq:Quantum2.5}) is exponentially suppressed. Therefore, this series generally converges rapidly, and permits accurate determination of the quantum solution $G(\mathbf{\hat r};E)$.  (This even includes tunneling sources with $E<0$ that have no classical counterpart.  The only exception is the plane $z=0$ where the series does not converge absolutely.  This is not surprising because $G(\mathbf{\hat r};E)$ itself diverges as $\mathbf{\hat r}$ approaches the origin, $G(\mathbf{\hat r};E) \sim - m / (2\pi\hbar^2 \hat r)$.)

\subsection{Currents generated by the source}
\label{subsec:Quantum3}

The stationary Schr\"odinger equation (\ref{eq:Quantum2.1}) contains an additional inhomogeneous term, the ``point source'' $C \delta^{(3)}(\mathbf{\hat r})$.  Defining the current density distribution in the scattering wave $\psi(\mathbf{\hat r}; E)$ in the usual manner,
\begin{equation}
\label{eq:Quantum3.1}
\mathbf j(\mathbf{\hat r};E) = \frac\hbar m \imag\bigl[\psi^*\boldsymbol\nabla \psi\bigr]
- \frac{q\mathbf A(\mathbf{\hat r})}m \bigl| \psi \bigr|^2 \;,
\end{equation}
we integrate $\mathbf j(\mathbf{\hat r};E)$ over a surface $\partial\Omega$ enclosing the origin and find the total current $J(E)$ emitted by the source:
\begin{equation}
\label{eq:Quantum3.2}
J(E) = \oint_{\partial\Omega} \mathbf j(\mathbf{\hat r};E) \cdot {\rm d}{\mathbf n}(\mathbf{\hat r}) = 
\int_\Omega {\rm d}^3\hat r \,\boldsymbol\nabla\cdot \mathbf j(\mathbf{\hat r};E) \;.
\end{equation}
Solutions of the usual homogeneous Schr\"odinger equation fulfill $\boldsymbol\nabla\cdot \mathbf j = 0$ and cannot carry a net flux through any closed surface.  The presence of the source modifies the equation of continuity, however \cite{Kramer2002a}:
\begin{equation}
\label{eq:Quantum3.3}
\boldsymbol\nabla\cdot \mathbf j(\mathbf{\hat r};E) = - \frac2\hbar 
\imag\bigl[C^* \delta^{(3)}(\mathbf{\hat r}) \psi(\mathbf{\hat r};E) \bigr] \;,
\end{equation}
and because $\psi(\mathbf{\hat r};E) = C G(\mathbf{\hat r};E)$, the total current is proportional to the imaginary part of the Green function at the source location \footnote{The imaginary part of the Green function $\psi = \imag[G(\mathbf{\hat r};E)]$ is a solution of the ordinary Schr\"odinger equation $[E - {\cal H}]\psi = 0$.  Unlike $G(\mathbf{\hat r};E)$ itself, it remains well-defined in the limit $\mathbf{\hat r} \rightarrow \mathbf 0$.}:
\begin{equation}
\label{eq:Quantum3.4}
J(E) = - \frac2\hbar |C|^2 \lim_{\mathbf{\hat r} \rightarrow \mathbf 0}  
\imag\bigl[G(\mathbf{\hat r};E) \bigr] \;.
\end{equation}

The parallel field Green function (\ref{eq:Quantum2.5}) yields a series representation for $J(E)$.  We obtain with $L_n(0) = 1$ \cite{Abramowitz1965a}:
\begin{equation}
\label{eq:Quantum3.5}
J(E) = - \frac{2m{\cal B}}{\beta\hbar^4 {\cal E}} |C|^2
\sum_{n=0}^\infty \Ai\bigl[-\beta (E - E_n) \bigr]^2 \;.
\end{equation}
$J(E)$ is experimentally accessible as the total $s$--wave photodetachment cross section.  Similar results for the $p$--wave cross section have been found using Fermi's Golden Rule \cite{Fabrikant1991a,Du1989c,Bivona1995a}.

\section{Examples}
\label{sec:Examples}

To illustrate the results of the two previous sections, we now plot and discuss radial profiles of the integrated density $n(\hat\rho) = 2\pi\hat\rho |\psi(\hat\rho)|^2$ for three cases that sample different regimes of motion in parallel fields.  (We use integrated densities because the electrons tend to cluster around the symmetry axis (see Fig.~\ref{fig:Primer0.1}); indeed, for $\hat\rho \rightarrow 0$ the semiclassical approximation $n_{\rm sc}(\hat\rho)$ approaches a constant if a focal line is present.)  All results shown are normalized to a total flux of one electron per second.

\subsection{Moderate Fields}
\label{subsec:Examples1}

\begin{figure}
\begin{center}
\includegraphics[width=\columnwidth]{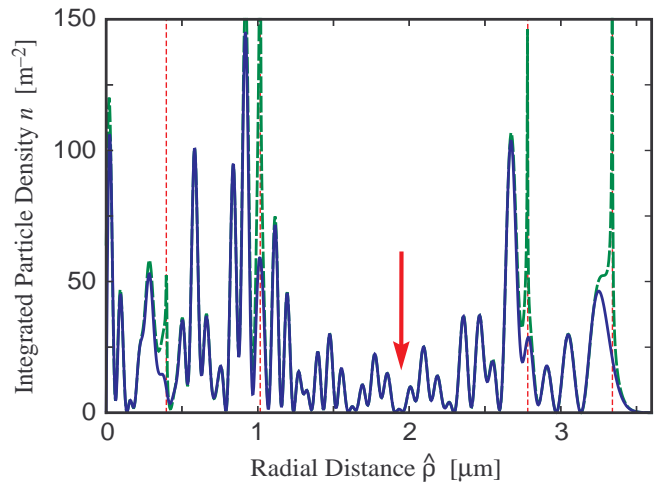}
\end{center}
\caption{Integrated density profile for a point source of electrons (normalized flux $1$~s$^{-1}$, energy $E = 10^{-4}$~eV) in parallel fields ${\cal E} = 15$~V/m, ${\cal B} = 0.02$~T ($\eta = 7.908$, $\epsilon = 86.38$), at a distance $\hat z = 30\,\mu$m from the source (cf.~Fig.~\protect{\ref{fig:Primer0.1}}).  Blue curve:  Quantum result (\protect{\ref{eq:Quantum2.5}}), green curve (dashed):  Semiclassical approximation (\protect{\ref{eq:Semi0.4a}}), including tunneling trajectories.  The WKB approximation diverges at the caustics (red dotted lines).  The red arrow marks the position of the ``snake'' closed orbit.
\label{fig:Examples1.1}
}
\end{figure}

We first return to the model problem used in the introduction (${\cal E} = 15$~V/m, ${\cal B} = 0.02$~T, $E = 10^{-4}$~eV, or $\eta = 7.908$, $\epsilon = 86.38$; see Fig.~\ref{fig:Primer0.1}) as an example with moderate parameters --- semiclassically, at most eight trajectories contribute, while the quantum mechanical solution converges after summation over about fifty Landau levels.  Figure~\ref{fig:Examples1.1} shows details of the radial density profile $n(\hat\rho)$ at a detector distance $\hat z = 30$~$\mu$m.  The semiclassical result $n_{\rm sc}(\hat\rho)$ (\ref{eq:Semi0.4a}) obtained including complex ``tunneling trajectories'' (green) is an excellent approximation to the exact solution $n(\hat\rho)$ (blue) given by the quantum Green function (\ref{eq:Quantum2.5}), except near the four intersections with caustic surfaces (red dotted lines).  The uniform expansion (\ref{eq:Semi3.1}) into Airy functions (not shown) is virtually indistinguishable from the exact result.  Its typical relative error is much smaller than 1~\%\ over most of the radial range, except in the immediate vicinity of the symmetry axis:  Near a focal line, the densities $\rho_{\rm sc}(\hat\rho)$ and $\rho_{\rm uni}(\hat\rho)$ diverge like $\hat\rho^{-1}$.  Therefore, $n_{\rm sc}(\hat\rho)$ and $n_{\rm uni}(\hat\rho)$ approach a constant value as $\hat\rho \rightarrow 0$, while $n(\hat\rho)$ vanishes linearly.  A proper treatment would require a different uniform approximation valid near focal line singularities.

Another feature of Fig.~\ref{fig:Examples1.1} is the ``metastructure'' visible in the interference pattern as a modulation in $n(\hat\rho)$ around $\hat\rho \sim 2\,\mu$m.  Its length scale ($\sim 0.3\,\mu$m) falls in between the period of individual oscillations in the density ($\sim 0.08\,\mu$m), and the range of the classical caustic structure ($\sim 1\,\mu$m).  An approximate symmetry of $n(\hat\rho)$ with respect to the location of the snake-type closed orbit (which intersects the detector plane at $\hat\rho_{\rm co} = 1.945\,\mu$m) is conspicuous.  We already pointed out the role of closed orbits in organizing the spatial interference pattern in the electronic density (see Fig.~\ref{fig:Primer0.1}) \footnote{Note that along a closed orbit, the phase difference of two interfering trajectories is fixed.}.

\subsection{Weak Magnetic Field}
\label{subsec:Examples2}

The addition of even a weak magnetic field to an electric field fundamentally alters the structure of the trajectories and the resulting interference pattern.  A pure electric field produces a ``fireworks'' structure of trajectories, and just two paths arrive at a point on the detector \cite{Bracher1998a,Kramer2002a}.  In contrast, with any magnetic field present, the emitted electrons initially spread outward, but the magnetic field eventually turns them around and refocuses them onto the symmetry axis, where they form a focal line segment that joins opposite cusps of adjacent caustic surfaces (Section~\ref{sec:Caustic}).  For small $\eta$, the caustics barely overlap and form a sequence of narrow, diamond-shaped \emph{focal regions} with four-path interference.  They are centered around $z_k = (k\pi)^2/\eta$ and have an approximate radial width $E/(q{\cal E})$.  Outside the focal regions, two-path interference prevails; the behavior matches the purely electric case, except that the cyclotron motion causes the radius of the pattern to oscillate.  With increasing distance $\hat z$, the radial electronic distribution expands and contracts, and the current is forced through a sequence of hourglass-shaped constrictions (see Fig.~\ref{fig:Caustic5.1}(a) and Ref.~\cite{Kramer2001a}).
\begin{figure}
\begin{center}
\includegraphics[width=\columnwidth]{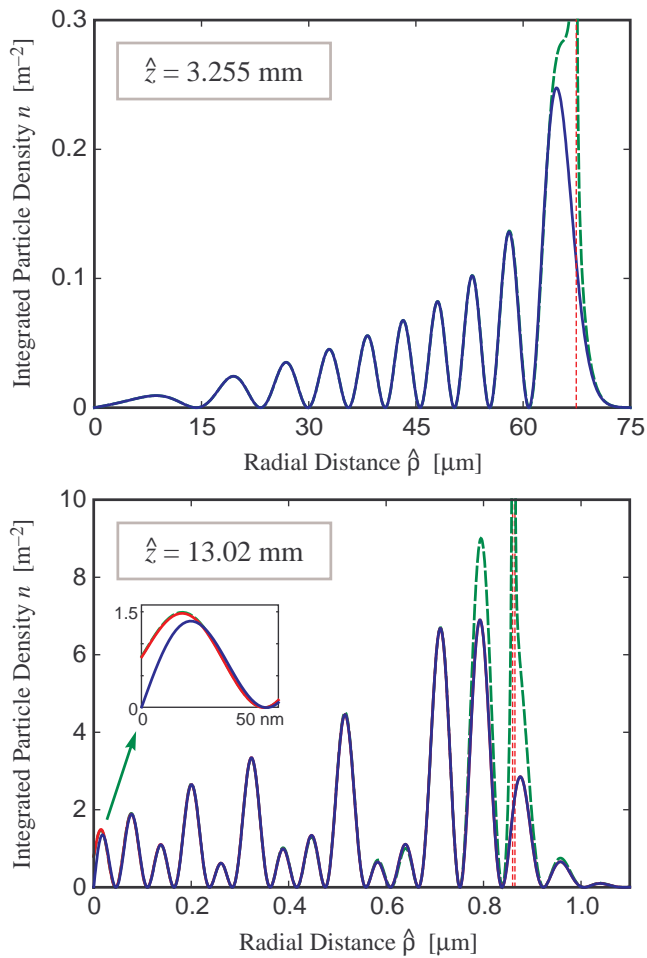}
\end{center}
\caption{Integrated density profiles for a normalized point source in fields ${\cal E} = 116$~V/m and ${\cal B} = 1$~mT (electron energy: $E = 10^{-4}$~eV, $\eta = 0.051$, $\epsilon = 1728$).  Top:  Distribution close to the maximum radial extension (detector distance $\hat z =3.255$~mm), bottom:  Distribution at the ``bottleneck'' constriction ($\hat z = 13.02$~mm; note the change in scale for the $\hat\rho$ axis).  The uniform approximation (\protect{\ref{eq:Semi3.1}}) (red) is virtually indistinguishable from the quantum result (\protect{\ref{eq:Quantum2.5}}) (blue), except near $\rho = 0$ (see detail). The semiclassical approximation including tunneling paths (\protect{\ref{eq:Semi0.4a}}) (dashed green curve) diverges at the caustics (red dotted lines). 
\label{fig:Examples2.1}
}
\end{figure}

Figure~\ref{fig:Examples2.1} depicts radial density profiles at maximum extension (top) and at the first ``bottleneck'' (bottom) for the set of parameters ${\cal E} = 116$~V/m, ${\cal B} = 1$~mT, and $E = 10^{-4}$~eV, or $\eta = 0.051$ and $\epsilon = 1728$.  In the first case ($\hat z = 3.255$~mm), we obtain a simple two-path interference pattern.  The semiclassical approximation (green) and the quantum solution (blue) agree very well except near the outer turning point at $\hat\rho = 67.4$~$\mu$m.  Because no focal line is present, the WKB approximation holds even for $\hat\rho=0$.  At the center of the constriction ($\hat z = 13.02$~mm), a more involved oscillation pattern emerges as four paths interfere with each other in the sector of classically allowed motion.  The semiclassical approximation fails near the double caustic at $0.862$~$\mu$m, and also close to the symmetry axis due to the presence of the focal line, but performs well otherwise.  To repair the divergence at the caustics we resort to the uniform approximation (red) which is indistinguishable from the quantum result except near $\rho = 0$.  Note that the oscillations in the density profile extend beyond the turning point; they indicate interference of the two tunneling trajectories.

For weak magnetic fields, a large number of terms ($\sim 10^3$) must be summed up to achieve convergence in the series for $G(\mathbf{\hat r};E)$ (\ref{eq:Quantum2.5}).  (Note that they cancel each other almost perfectly everywhere outside the classically allowed region; this cancellation over a large region is especially remarkable near the bottleneck.)  In contrast, the semiclassical method relies on at most four trajectories.  Here, the uniform approximation (\ref{eq:Semi3.1})--(\ref{eq:Semi3.3b}) is a very efficient scheme to evaluate the wave function.

\subsection{Strong Magnetic Field}
\label{subsec:Examples3}

\begin{figure}
\begin{center}
\includegraphics[width=\columnwidth]{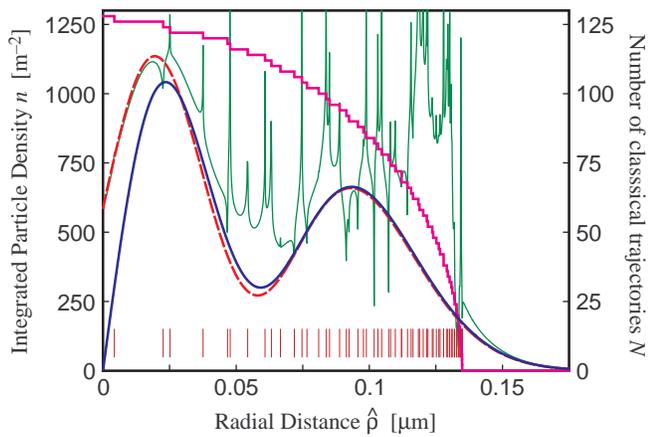}
\end{center}
\caption{Integrated density profile (normalized flux) for ${\cal E} = 15$~V/m, ${\cal B} = 0.5$~T, and $E = 10^{-4}$~eV ($\eta = 197.7$, $\epsilon = 3.455$), at a detector distance $\hat z =30\,\mu$m.  Blue curve (bold):  Quantum result (\protect{\ref{eq:Quantum2.5}}), red curve (dashed):  Uniform approximation (\protect{\ref{eq:Semi3.1}}),  green line (thin): WKB approximation (\protect{\ref{eq:Semi0.4a}}) with tunneling contributions.  The sequence of bars (red) marks the position of the 64 caustics.  The purple curve indicates the number of interfering classical trajectories $N(\hat\rho)$.
\label{fig:Examples3.1}
}
\end{figure}

Finally, we take a look at the opposite limit of a strong magnetic field $\cal B$.  Figure~\ref{fig:Examples3.1} displays the density profile at a distance $\hat z = 30$~$\mu$m for  ${\cal E} = 15$~V/m, ${\cal B} = 0.5$~T, and $E = 10^{-4}$~eV, corresponding to dimensionless variables $\eta = 197.7$ and $\epsilon = 3.455$.  The situation is most naturally described in terms of the quantum result $n(\hat\rho)$ (blue curve) --- effectively, only two open ``channels'' contribute to the series for $G(\mathbf{\hat r}';E)$ (\ref{eq:Quantum2.5}).  In contrast, up to 128 classical trajectories connect the source to a point on the detector.  The 64 intersections with caustics (marked by red bars) are spread over the entire range of classical motion, and they are most densely clustered near its outer limit $\rho = 1$ (or  $\hat\rho = 0.135$~$\mu$m).  These caustics render the primitive semiclassical result (\ref{eq:Semi0.4a}) virtually worthless.  However, it is striking that the uniform semiclassical approximation $n_{\rm uni}(\hat\rho)$ (\ref{eq:Semi3.1}) still performs well over most of the range of $\hat\rho$ (except near $\hat\rho = 0$, where it fails due to the presence of focal lines).  This is an example where we may reasonably expect this approximation to fail, but it gives a surprisingly good result.  We conclude that the expansion into Airy functions (dashed red curve) is a reliable approximation over a wide range of parameters.

\section{Conclusion and Outlook}
\label{sec:Outlook}

We studied the dynamics of monoenergetic electrons emitted from a pointlike, isotropic source placed in homogeneous, parallel electric and magnetic fields $\boldsymbol{\cal E}$, $\boldsymbol{\cal B}$.  The problem is simple enough to permit a full quantum solution in the form of a series for the Green function, Eq.~(\ref{eq:Quantum2.5});  likewise, analytic expressions are available for the classical electron motion, and we used them to develop semiclassical and uniform Airy approximations to the wave function.

By tuning the two ``control parameters'' $\eta = v_0{\cal B}/{\cal E}$ and $\epsilon = E/(\hbar\omega_L)$ that characterize the system in the classical and quantum domain, both the number of interfering classical trajectories and the number of open quantum scattering channels can be adjusted independently.  This is accomplished by simply varying the electron energy and/or the external fields. 

We verified that the patterns seen in the quantum density distribution are naturally explained in terms of the caustic surfaces and closed orbits of the corresponding classical motion.  Using a temporal parametrization technique, we examined the set of caustic surfaces and found a surprisingly complex classification scheme featuring seven distinct types of caustics.  With its help, we learned how the caustic pattern systematically evolves with increasing $\eta$.

The present state of our study suggests future work in two directions:  While we achieved a complete description of the caustic structure for $\rho > 0$, we did not examine in detail the wave functions near the focal lines, cusps, and umbilical points on the $\rho = 0$ axis.  In the future we plan a systematic study of these catastrophes and their proper uniform approximations in the parallel field problem.

Pointlike sources of electrons are almost perfectly realized in near-threshold photodetachment of negative atomic ions.  Photodetachment experiments in parallel fields should therefore provide practical access to a system where the interference of matter waves can be precisely tailored.  For the field strengths typically used in photodetachment microscopy, the interference patterns tend to be very small, however (with diameters in the $\mu$m range), which makes them hard to observe.  To overcome this difficulty, we will examine non-uniform field configurations that ``flare out'' from the source and effectively magnify the density distribution on the detector.  Given the excellent accuracy of the semiclassical and uniform approximations reported in this paper, we expect that semiclassical models likewise yield good predictions for the interference patterns in these more sophisticated field geometries.

\acknowledgments This work was supported by the National Science Foundation (CB and JD) and the DFG, Emmy Noether Programm KR 2889/1-1 (TK).

\appendix

\section{Temporal parametrization of caustic surfaces}
\label{sec:Appendix}

In Section~\ref{sec:Caustic}, we described the caustic structure emerging in parallel fields, and its evolution with changing value of the parameter $\eta$.  The results presented there were based on the temporal parametrization $\bigl[ \rho_\pm(t), z_\pm(t) \bigr]$ (\ref{eq:Caustic1.3}) of the caustic surfaces.  However, for the sake of clarity, we furnished no proof in the main text.  We now discuss the mathematical properties of the parametrization and show how they may be exploited for a complete characterization of the caustics.  While our argument employs only elementary mathematics, the complete description is surprisingly complex.

\subsection{Proof of the algorithm in Section~\ref{subsec:Caustic2}}
\label{subsec:Algorithm}

We first give a proof for the algorithm shown in Figure~\ref{flowchart}.  The existence of solutions in (\ref{eq:Caustic1.3}) is contingent upon the conditions laid out in Eqs.~(\ref{eq:Caustic2.2a}), (\ref{eq:Caustic2.2b}).  We consider only $t>0$.

\subsubsection{$\tau < 0$}

We start with the case $\tau = \tan t < 0$.  Then, $A(\tau,t) > \tau^2 > 0$ (\ref{eq:Caustic1.3a}) is always positive, and we can restrict our attention to the second inequality.

Since $t + \sqrt{A} > t + |\tau|$, the second condition in (\ref{eq:Caustic2.2b}) is always satisfied for the $(+)$ solution.  We show next that Eq.~(\ref{eq:Caustic2.2b}) is fulfilled for the $(-)$ solution only if $t \leq \frac\eta2$ holds.  For the proof, we compare both sides in this inequality:
\begin{equation}
\label{eq:Algo1}
(t-\tau)^2 \leq \bigl(t - \sqrt{A} \bigr)^2 \;.
\end{equation}
When multiplied out, this is equivalent to:
\begin{equation}
\label{eq:Algo2}
2t\sqrt{A} \leq \Bigl(2t - \frac{\eta^2}t \Bigr)\tau + \eta^2 = B(\tau,t) \;.
\end{equation}
To facilitate comparison, we first square both sides and form their difference:
\begin{equation}
\label{eq:Algo3}
\Delta(\tau,t) = B^2 - 4t^2 A = \eta^2 \bigl(\eta^2 - 4t^2 \bigr) \Bigl(1 - \frac\tau t \Bigr)^2 \;,
\end{equation}
from which it follows (barring $\tau = t$) that $\mathrm{sgn}\,\Delta(\tau, t) = \mathrm{sgn}(\frac\eta2 - t)$.  The condition (\ref{eq:Algo2}) is then equivalent to $\Delta \geq 0$ \textit{and} $B \geq 0$.  In the interval $t > \frac\eta2$, the first inequality is never satisfied, and hence the $(-)$ solution is complex there.  For $t \leq \frac\eta2$ and $\tau < 0$, however, both inequalities hold, and the $(-)$ solution is real.

\subsubsection{$\tau > 0$}

Now we assume $\tau > 0$.  Using the same line of argument, we show that for the $(-)$ solution, Eq.~(\ref{eq:Caustic2.2a}) implies $2t\sqrt{A} \geq B$, or equivalently, $\Delta \leq 0$ \textit{or} $B \leq 0$.  Therefore, the condition is fulfilled if $t \geq \frac\eta2$.  Since $A(\tau,t) \geq 0$ always holds in this regime, we find that the $(-)$ solution is real for $t>\frac\eta2$ and $\tan t > 0$.  For the $(+)$ solution, we find the condition $-2t\sqrt{A} \geq B$, which is identical to the inequality pair $\Delta \geq 0$ \textit{and} $B \leq 0$.  For $t > \frac\eta2$, they are never simultaneously fulfilled, so the $(+)$ solution is complex for $\tau > 0$ \footnote{This includes the case $\tau = \tan t = t$, where $\Delta(t,t) = 0$, but $B(t,t) = 2t^2 > 0$.}.

\begin{figure}
\begin{center}
\includegraphics[width=\columnwidth]{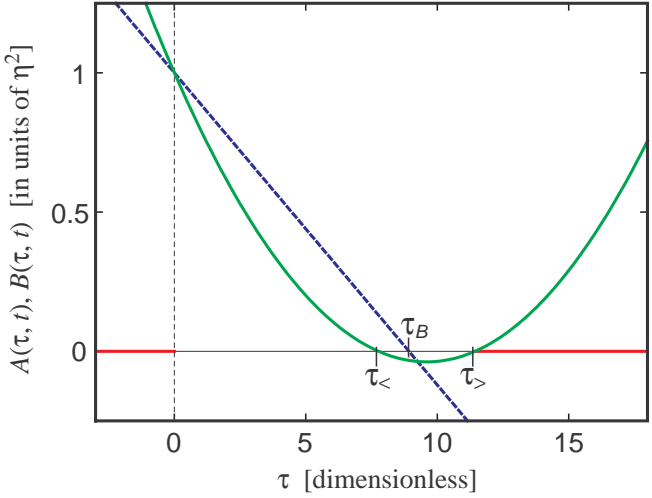}
\end{center}
\caption{The auxiliary functions $A(\tau,t)$ (\protect{\ref{eq:Caustic1.3a}}) (green, solid curve) and $B(\tau,t)$ (\protect{\ref{eq:Algo2}}) (blue, dotted line) for $t \leq \frac\eta2$.  In the red intervals, the parametrization (\ref{eq:Caustic1.3}) describes a pair of physical solutions.
\label{fig:Algo1}
}
\end{figure}

However, these inequalities show that for $\tau > 0$, both solutions will be real in the interval $t \leq \frac\eta2$ if $A(\tau,t) \geq 0$ \textit{and} $B(\tau,t) \leq 0$, and otherwise complex.  $A(\tau,t)$ is a quadratic function of $\tau$ that formally has roots:
\begin{equation}
\label{eq:Algo4}
\tau_{\gtrless} = \frac{\eta^2}{2t} \left(1 \pm \sqrt{1 - \frac{4t^2}{\eta^2}} \right) \;.
\end{equation}
Within the interval $\tau_< < \tau < \tau_>$, $A(\tau, t)$ is negative.  Similarly, $B(\tau,t)$ is a linear function of $\tau$ with negative slope; $B(0,t) = \eta^2$, and $B(\tau,t)$ remains positive for $0 \leq \tau < \tau_B = t\eta^2 / (\eta^2 - 2t^2)$.  In both intervals, all solutions are complex.  At the zero $\tau_B$ of $B(\tau,t)$, however, $A(\tau_B,t)$ is negative:
\begin{equation}
\label{eq:Algo5}
A(\tau_B,t) = \tau_B^2 \Bigl(\frac{4t^2}{\eta^2} -1 \Bigr) \;.
\end{equation}
Therefore, the conditions $A(\tau,t) \geq 0$ \textsl{and} $B(\tau, t) \leq 0$ are simultaneously satisfied when, and only when, $\tau \geq \tau_>$ (see Figure~\ref{fig:Algo1}).

\subsubsection{$t = \eta/2$}

We finally examine the behavior at the boundary $t = \frac\eta2$.  Above, we inferred that the solution space is degenerate if \textsl{either} ($\tau = \tan \frac\eta2 < 0$), \textsl{or} ($A(\tau,t) > 0$ \textsl{and} $B(\tau, t) < 0$).  Because
\begin{equation}
\label{eq:Algo6}
A(\tau,t) = \Bigl(\eta - \tan\frac\eta2 \Bigr)^2 \;, \quad
B(\tau,t) = \eta \Bigl(\eta - \tan\frac\eta2 \Bigr) \;,
\end{equation}
the latter condition holds for $\tan\frac\eta2 > \eta$.  In the intervals $0 < \tan\frac\eta2 < \eta$, only the $(-)$ solution is real.

\subsection{Irregular points on caustics}
\label{subsec:Caustic3App}

\begin{table*}
\begin{tabular*}{\textwidth}{@{\extracolsep{\fill}}|c|c|c|c|m{0.35\textwidth}|}
\hline
\multicolumn{5}{|c|}{Irregular Points for General Values of $\eta$}\\
\hline
\hline
$t_0$ & $z_0$ & $\rho_0$ & Local Behavior of Caustics & Remarks\\ 
\hline
$t_0 = (k - \frac12)\pi$ &
$\begin{gathered}
-\frac{t_0^2}\eta
\end{gathered}$ &
$1$ &
$\begin{gathered}
\delta z \sim \frac{t_0(\eta^2 + 4)}{2\eta}\,\delta t \\
\delta\rho \sim -\frac{\eta^2}{2t_0^2(\eta^2 + 4)}\,\delta z^2\\
\end{gathered}$ &
Maximum radial extension of caustics; singularity in the parametrization only; switch from ($+$) solution for $\delta t > 0$ to ($-$) solution for $\delta t < 0$ \\
\hline
$\begin{gathered}
\textstyle t_0 = (k - \frac12)\pi \\
\textstyle \eta > (2k-1)\pi
\end{gathered}$ &
$\begin{gathered}
\frac{t_0^2}\eta
\end{gathered}$ &
$\begin{gathered}
\frac1\eta \sqrt{\eta^2 - 4t_0^2}
\end{gathered}$ &
$\begin{gathered}
\delta z \sim -\frac{t_0 [\eta^2 + 4(1-t_0^2)]}{2\eta}\,\delta t \\
\delta\rho \sim \frac{2}{\eta\rho_0}\,\delta z\\
\end{gathered}$ &
Singularity in the parametrization, existing only for $t_0 < \frac\eta2$; switch from ($-$) solution for $\delta t > 0$ to ($+$) solution for $\delta t < 0$.  Corresponds to a ``snake''--type closed orbit\\
\hline
$\begin{gathered}
\textstyle t_0 = \frac\eta2 \\
\textstyle \eta \neq k\pi,\\
\textstyle \eta \neq \eta_{\rm crit}^{k}
\end{gathered}$ &
$\begin{gathered}
-\frac\eta4
\end{gathered}$ &
$0$ &
$\begin{gathered}
\delta z \sim \frac{\tan\frac\eta2}{\eta - \tan\frac\eta2}\,\delta t \\
\delta\rho^2 \sim \frac4\eta\, \sin^2\frac\eta2\,\delta z\\
\end{gathered}$ &
Uphill turning point of parabolic shape; singularity of the parametrization only.  $\delta t$ is positive for $2(k-1)\pi < \eta < \eta_{\rm crit}^k$, and negative for $\eta_{\rm crit}^k < \eta < 2k\pi$\\
\hline
$\begin{gathered}
\textstyle t_0 = t_{\min}^{(k)} \\
\textstyle \eta > \eta_{\rm crit}^{k}
\end{gathered}$ &
$\begin{gathered}
-\frac{\eta t_0^2}{\tan^2 t_0}
\end{gathered}$ &
$\sqrt{\sin^2 t_0 - \eta^2 \cos^2 t_0}$ &
$\begin{gathered}
\delta z^2 \sim \frac{\eta^2 t_0^2}{\tan^4 t_0}\,
   \left.\frac{\rmd A(t)}{\rmd t}\right|_{t_0}\,\delta t \\
\delta\rho \sim \frac{\eta\sin(2t_0)}{2\rho_0 t_0}\,\delta z\\
\end{gathered}$ &
Minimum time of flight along caustic; singularity of the parametrization only.  Switch from ($+$) solution for $\delta z < 0$ to ($-$) solution for $\delta z > 0$\\
\hline
$t_0 = k\pi$ & 
$\begin{gathered}
\frac{k\pi(k\pi + \eta)}\eta
\end{gathered}$ &
$0$ &
$\begin{gathered}
\delta z \sim \frac{3(\eta + 2k\pi)}{2\eta}\,\delta t \\
\delta\rho^2 \sim -\frac{8\eta^2}{27k\pi(\eta + 2k\pi)^2}\,\delta z^3\\
\end{gathered}$ &
Cylindrical cusp, opening towards the source ($\delta z < 0$).  Physical singularity of the ($+$) solution. $\delta t$ must be negative \\ 
\hline
$\begin{gathered}
t_0 = k\pi \\ 
\eta \neq 2k\pi
\end{gathered}$ &
$\begin{gathered}
\frac{k\pi(k\pi - \eta)}\eta
\end{gathered}$ &
$0$ &
$\begin{gathered}
\delta z \sim -\frac{3(\eta - 2k\pi)}{2\eta}\,\delta t \\
\delta\rho^2 \sim \frac{8\eta^2}{27k\pi(\eta - 2k\pi)^2}\,\delta z^3\\
\end{gathered}$ &
Cylindrical cusp, opening in field direction ($\delta z > 0$).  Physical singularity of the ($-$) solution. Sign of $\delta t$ depends on $t_0$: $\delta t > 0$ for $t_0 > \frac\eta2$, $\delta t < 0$ for $t_0 < \frac\eta2$ \\ 
\hline
\multicolumn{5}{c}{\ }\\
\hline
\multicolumn{5}{|c|}{Irregular Points for Special Values of $\eta$}\\
\hline
\hline
$t_0$ & $z_0$ & $\rho_0$ & Local Behavior of Caustics & Remarks\\ 
\hline
$\begin{gathered}
\textstyle t_0 = k\pi \\
\textstyle \eta = 2k\pi
\end{gathered}$ &
$\begin{gathered}
-\frac\eta4
\end{gathered}$ &
$0$ &
$\begin{gathered}
\delta z \sim \frac1{k\pi}\,\delta t^2 \\
\delta\rho \sim \delta z\\
\end{gathered}$ &
Umbilical point in the ($-$) solution; physical singularity of the caustic set; two surfaces (distinguished by the sign of $\delta t$) form a double cylindrical cone with opening angle $\frac\pi4$\\
\hline
$\begin{gathered}
\textstyle t_0 = \frac\eta2 \\
\textstyle \eta = \eta_{\rm crit}^{k}
\end{gathered}$ &
$\begin{gathered}
-\frac\eta4
\end{gathered}$ &
$0$ &
$\begin{gathered}
\delta z^2 \sim \eta \delta t \\
\delta\rho^2 \sim \frac{4\eta^2}{1 + \eta^2}\,\delta z\\
\end{gathered}$ &
Bifurcation point in the temporal parametrization (caustic remains parabolic); solution at top of caustic switches to ($+$) branch; extreme sensitivity to changes in $\delta t \geq 0$\\
\hline
$\begin{gathered}
\textstyle t_0 = (k - \frac12)\pi \\
\textstyle \eta = (2k-1)\pi
\end{gathered}$ &
$\begin{gathered}
-\frac\eta4
\end{gathered}$ &
$0$ &
$\begin{gathered}
\delta z \sim -\delta t \\
\delta\rho^2 \sim \frac4\eta\, \delta z\\
\end{gathered}$ &
Creation of a ``snake''--type closed-orbit; singularity of the parametrization only.  
At the top of caustic (which remains parabolic), solution switches to ($-$) branch\\
\hline
\end{tabular*}
\caption{Irregular points of the temporal parametrization $\bigl[z_\pm(t),\rho_\pm(t)\bigr]$ (\protect{\ref{eq:Caustic1.3}}), together with local leading-order expansions for the caustic surfaces.  (The minimal times $t_{\min}^{(k)}$ and critical parameter values $\eta_{\rm crit}^k$ are defined in Eqs.~(\protect{\ref{eq:Caustic2.2d}}) and (\protect{\ref{eq:Caustic3.0a}}).)
\label{tab:Caustic3.1}}
\end{table*}
A point on the caustic $(t_0,z_0,\rho_0)$ is \emph{regular} if the functions $\rho_\pm(t)$, $z_\pm(t)$, and either $\rho(z)$ or $z(\rho)$ have continuous derivatives ($C^\infty$) at this point.  Otherwise the point is \emph{irregular}.  The caustic surfaces in space must be smooth unless one or both of the functions $\rho_\pm(t)$, $z_\pm(t)$ become irregular, or both derivatives ${\rm d}\rho_\pm(t)/{\rm d}t$ and ${\rm d}z_\pm(t)/{\rm d}t$ simultaneously vanish.

The parametrization (\ref{eq:Caustic1.3}) has irregular points at (a), (b) $t_0 = (k - \frac12)\pi$ so $\tau =\tan t$ diverges; (c) $t_0 = \frac\eta2$ so the radicand in $\rho_\pm(t)$ vanishes for either the ($+$) or the ($-$) solution; (d) the minimal values $t_0 = t_{\min}^{(k)}$ for which $A(\tan t_0,t_0) = 0$ (\ref{eq:Caustic2.2d}) and $B(\tan t_0, t_0) \leq 0$ (see above); (e) $t_0 = k\pi$ so $\tau = 0$ (see Section~\ref{subsec:Caustic3}).  These irregular points occur for general values of $\eta$, and each case occurs at most once in every cyclotron period $(k-1)\pi < t \leq k\pi$.  For isolated values of $\eta$, two of these conditions may coincide.  We will address this issue separately below.

The behavior of the caustic surface and its temporal parametrization near the irregular points $(t_0,z_0,\rho_0)$ may be inferred from a local expansion of Eq.~(\ref{eq:Caustic1.3}) into powers of the deviations $\delta t = t-t_0$, $\delta\rho =  \rho -\rho_0$, $\delta z = z-z_0$ \footnote{Alternatively, one might use the implicit pair of equations (\protect{\ref{eq:Class2.12}}) and (\protect{\ref{eq:Caustic1.2}}) for this purpose.}.  The results of this tedious yet straightforward calculation are assembled in Table~\ref{tab:Caustic3.1}.  Here, we proceed with an interpretation of our findings.

\subsubsection{Maxima of $\rho$: $t_0 = (k - \frac12)\pi > \frac\eta2$}
\label{subsubsec:Caustic3.1}

This is the simplest case.  Here, $\tau = \tan t$ diverges, and as $t$ increases through this point we switch from the $(-)$ solution to the $(+)$ solution (see algorithm in Fig.~\ref{flowchart}).  The caustic is a smooth curve in $(\rho,z)$ space.  This point represents the maximum radial extent of the caustics.  

\subsubsection{Maxima of $\rho$ vs.\ endpoints of closed orbits: $t_0 = (k - \frac12)\pi < \frac\eta2$}
\label{subsubsec:Caustic3.2}

Here, the caustics are smooth curves in ($\rho,z$) space but the parametrization is tricky.  Both $(+)$ and $(-)$ solutions exist, but are discontinuous:  Passing through the divergence of $\tan t$, the $(-)$ curve $\bigl[ \rho_-(t),z_-(t) \bigr]$ joins smoothly to the $(+)$ curve $\bigl[ \rho_+(t),z_+(t) \bigr]$, and vice versa.  As $\delta t$ increases continuously along the caustic, if we switch from the $(-)$ solution to the $(+)$ solution, the caustic has a radial maximum, just as in case (a).  If we switch from the $(+)$ solution to the $(-)$ solution, the caustic curve is again smooth, and the irregular point represents the endpoint of a ``snake''--type closed orbit where $\dot\rho = \dot z = 0$.

\subsubsection{Top of caustic: $t_0 = \frac\eta2$, $\eta \neq k\pi$ and $\eta \neq \eta_{\rm crit}^k$}
\label{subsubsec:Caustic3.3}

This is the topmost point of the caustics ($z_0 = -\frac\eta4$), where $z(\rho)$ has a quadratic maximum.  The parametrization is again tricky:  $\delta z$ is proportional to $\delta t$, and $\delta \rho$ is proportional to $|\delta t|^{1/2}$.  This irregular point in the parametrization occurs in the $(-)$ solution if $\tan\frac\eta2 < \eta$, and in the $(+)$ branch otherwise.  Further analysis (see Table~\ref{tab:Caustic3.1}) shows that $\delta t$ must be positive if $2(k-1)\pi < \eta < \eta_{\rm crit}^k$, and negative if $\eta_{\rm crit}^k < \eta < 2k\pi$.  The critical values of $\eta$ are given by $\eta_{\rm crit}^k = \tan\bigl(\eta_{\rm crit}^k / 2 \bigr)$ (\ref{eq:Caustic3.0a}).

\subsubsection{Minimal times: $A(\tan t_0,t_0) = 0$}
\label{subsubsec:Caustic3.4}

Again, the caustic curve is smooth but the parametrization is irregular.  If $t<\frac\eta2$ and $\eta > \eta_{\rm crit}^k$ then in each cyclotron period $(k-1)\pi < t < k\pi$ there are two times at which $A(\tan t,t) = 0$, but in the vicinity of the smaller root, no physical solution exists for other reasons (see previous section).  Therefore, only the larger of the two roots, the minimal time $t_{\rm min}^{(k)}$ (\ref{eq:Caustic2.2d}), is an irregular point.  As $t$ increases through $t_{\rm min}^{(k)}$, the radicand $A(\tau,t)$ (\ref{eq:Caustic1.3a}) passes from negative to positive, and a pair of solutions $\bigl[z_\pm(t), \rho_\pm(t) \bigr]$ springs into existence.  The time $t_0$ hence marks a local minimum of the time of flight from the source to the caustic.  From this point, the caustic is described by the ($-$) solution for positive $\delta z$, and by the ($+$) solution for $\delta z < 0$.  In the vicinity of the singularity, small variations $\delta t \geq 0$ of the time of flight cause large shifts in position.

\subsubsection{Cusps: $t_0 = k\pi, \eta \neq 2k\pi$}
\label{subsubsec:Caustic3.5}

\begin{figure}
\begin{center}
\includegraphics[width=\columnwidth]{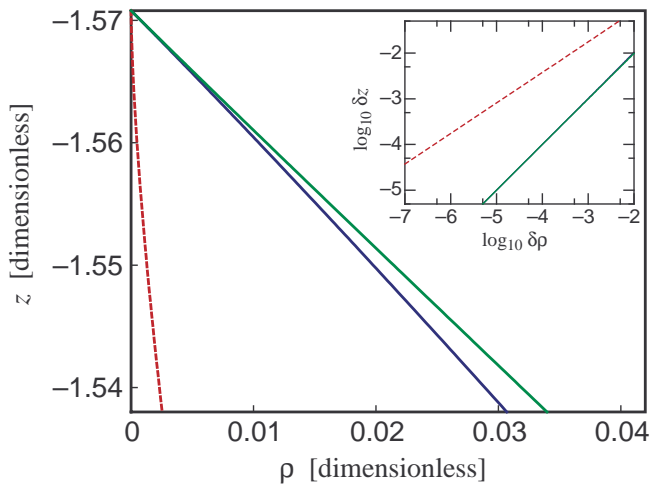}
\caption{Cusps and cones in caustic surfaces near the irregular point ($\rho_0 = 0, z_0 = -\frac\pi2$).  In the umbilic case $t_0 = \pi$ and $\eta=2\pi$, a pair of surfaces asymptotically coalesces into a double cone of opening angle 45$^\circ$ (``early'' caustic $\delta t < 0$, green line; ``late'' caustic $\delta t > 0$, blue curve). In contrast, for $t_0 = 2\pi$ the generic value $\eta = 8\pi/3$ leads to a cylindrical cusp \cite{Peters1994a} (red dotted curve, $\delta t > 0$).  A log-log plot of $\delta z$ versus $\delta\rho$ (inset) illustrates the different power laws $\delta\rho \sim \delta z$ and $\delta\rho^2 \sim C(\eta)\delta z^3$ guiding these surfaces, respectively.
\label{fig:Caustic3.1}}
\end{center}
\end{figure}
In all previous cases, the irregularity occurs in the time parametrization but the caustic itself is a smooth curve in $(\rho,z)$ space.  This case is different.  When $t_0 = k\pi$, $\tau$ vanishes, and the irregular points lie on the $z$--axis at $z_0^{\pm} = \frac{k\pi}\eta(k\pi \pm \eta)$.  As stated in Table~\ref{tab:Caustic3.1}, in the vicinity of these points the relationship between $\delta z$ and $\delta t$ is linear, while a more complex law connects $\delta\rho$ and $\delta t$:
\begin{equation}
\label{eq:Caustic3.0b}
\delta\rho_\pm^2 \sim -\frac{\eta \pm 2k\pi}{\eta k\pi} \delta t^3 \;.
\end{equation}
For a meaningful result, the sign of $\delta t$ in Eq.~(\ref{eq:Caustic3.0b}) must be properly chosen: We always have $\delta t < 0$ in the $(+)$ solution; in the $(-)$ solution, $\delta t$ is positive (negative) for $t_0 > \frac\eta2$ ($t_0 < \frac\eta2$).  (The special case $\eta = 2k\pi$ will be discussed later.)  Eliminating $\delta t$, we find that the caustic surface is locally parametrized in the form $\delta\rho_\pm^2 \pm C(\eta) \delta z_\pm^3 = 0$ where $C(\eta) > 0$ is a positive coefficient.  This is the characteristic functional form of a \emph{cusp singularity} \cite{Poston1978a}.  The caustic surface follows from rotation of the cusp profile around the symmetry axis and has been called a \emph{cylindrical cusp} \cite{Peters1997b}.

The $(+)$ and $(-)$ solutions produce cusps that point in opposite directions:  The cusp in the $(+)$ solution points downhill while that in the $(-)$ solution points uphill in Fig.~\ref{fig:Caustic4.1}.  Between each pair of cusps at $z_0^\pm(k\pi)$ a \emph{focal line} segment extends along the $z$--axis, where trajectories from all values of the polar angle $\phi'$ are focused onto $\rho = 0$.  These trajectories simultaneously arrive at the $z$--axis after $k$ complete cyclotron orbits.  Their emission angles $\theta'$ vary between 0 (downhill) for $z_0^+$ and $\pi$ (uphill) for $z_0^-$.  Depending on $k$ and $\eta$, the cusps at $z_0^+$ and $z_0^-$ may or may not belong to the same caustic surface.

\subsection{Special irregular points of caustics}
\label{subsec:Caustic3aApp}

\begin{figure}
\begin{center}
\includegraphics[width=\columnwidth]{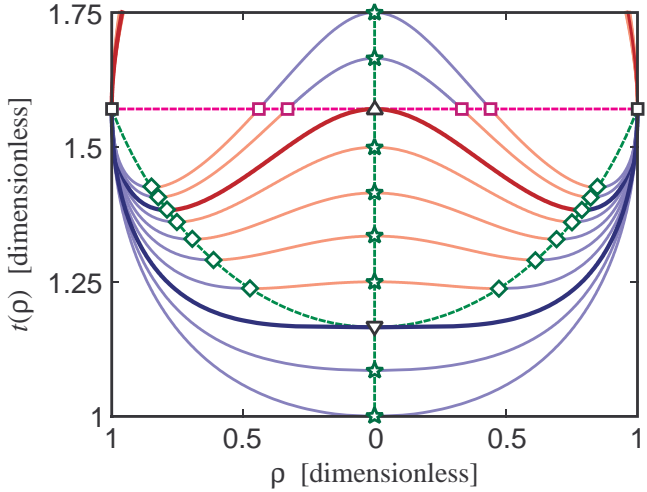}
\caption{Bifurcation and creation of a ``snake''--type closed orbit in the first caustic surface ($k=1$).  The figure displays a family of graphs of the function $t(\rho)$ with increasing values of $\eta$ (bottom to top):  For $\eta \leq \eta_{\rm crit}^1$, $t(\rho)$ has its minimum on the symmetry axis at the top of the caustic, $t(0)=\frac\eta2$ (case 2c), and $\rho(t)$ grows monotonically in $0 \leq \rho < 1$ (bottom curves, $\eta = 2,\,2.17$, corresponding to the late variant [Sl] of the ``teardrop'' caustic.  See also Fig.~\protect{\ref{fig:Caustic4.1}}.)  The behavior changes at the critical value $\eta_{\rm crit}^1 = 2.33112...$ (bottom heavy curve):  For $\eta > \eta_{\rm crit}^1$, $\rho=0$ is a local maximum of $t(\rho)$, while the minimum $t = t_{\rm min}^{(1)}$ (case 2d) is now attained off-axis.  We are at the intermediate-time variant [Si] of the ``teardrop'' ($\eta = 2.50, \,2.67, \,2.83, \,3$).  The extrema of $t(\rho)$ are located on the dashed green curve (``pitchfork bifurcation''). --- At $\eta=\pi$ (top heavy curve), the first snake orbit with time of flight $t = \frac\pi2$ (dashed purple line) detaches from the symmetry axis \protect{\cite{Peters1994a}}, and we enter the early regime [Se] of the caustic ($\eta = 3.33,\,3.5$, top curves). --- Red and blue segments represent the ($+$) and ($-$) solution in Eq.~(\protect\ref{eq:Caustic1.3}), respectively.  Symbols indicate irregular points of the parametrization: Half-integer orbits ($\Box$, cases 2a, 2b), the top of the caustic ($\boldsymbol\star$, case 2c), the minimal time $t_{\rm min}^{(1)}$ ($\Diamond$, case 2d), the bifurcation point ($\bigtriangledown$, case 3b), and creation of a closed orbit ($\bigtriangleup$, case 3c).
\label{fig:Caustic3.2}}
\end{center}
\end{figure}
The above irregular points are ``generic'' in the sense that they occur at almost all values of $\eta$.  Three extra cases arise only for special values of $\eta$, when two generic irregular points coalesce.

\subsubsection{Cylindrical hyperbolic umbilic: $t_0 = \frac\eta2 = k\pi$}
\label{subsubsec:Caustic3a.1}

If $\eta$ is an integer multiple of $2\pi$, then the time to arrive at the top of the caustic, $t_0 = \frac\eta2$ (case 2c), coincides with a completed cyclotron orbit (case 2d).  As $\eta$ increases to $2k\pi$, a cusp moves up to collide with the top of the caustic.  The cylindrical cusp and the smooth dome are changed into a double cylindrical cone having two branches, both meeting the $z$--axis at an angle of $\frac\pi4$.  These branches are both described by the ($-$) solution which is continuous at $t_0 = k\pi$; they differ in the sign of $\delta t$ (note that $\delta\rho \sim \delta z \sim \delta t^2/k\pi$).  Figure~\ref{fig:Caustic3.1} shows the caustic for $\eta = 2\pi$.  In Catastrophe Theory \cite{Thom1975a,Poston1978a,Berry1976a,Connor1976a}, this represents the generic catastrophe in three control parameters ($\rho,z,\eta$) known as the \emph{hyperbolic umbilic}.

\subsubsection{Bifurcation points of the parametrization: $t_0 = \frac\eta2, \eta = \eta_{\rm crit}^k$}
\label{subsubsec:Caustic3a.2}

This irregular point occurs at the critical values of $\eta$ given by Eq.~(\ref{eq:Caustic3.0a}) when cases (2c) and (2d) coincide.  As $\eta$ increases through $\eta_{\rm crit}^k$, keeping always $t_0 = \frac\eta2$, then according to the algorithm in Fig.~\ref{flowchart}, the $(-)$ solution continues, but the $(+)$ solution suddenly becomes real: In the process, a pair of real roots of $A(\tan t_0, t_0) = 0$ is created in the $k$th cyclotron period, and the larger of these roots $t_{\rm min}^{(k)}$ presents the minimal time of flight along the caustic (case 2d).  Concurrently, the top of the caustic ($t_0 = \frac\eta2$, case 2c) changes from a local \emph{minimum} of $t$ into a local \emph{maximum} (see Table~\ref{tab:Caustic3.1}).  

For $k=1$, the transition is illustrated in Fig.~\ref{fig:Caustic3.2}: Instead of the possibly double-valued function $\rho_\pm(t)$, we examine the inverted function $t(\rho)$ which is locally smooth (we limit ourselves to the top half of the uppermost smooth dome caustic).  We now look at the position of the extrema of $t(\rho)$ for varying values of $\eta$ (green dotted line) that correspond to the singularities (branching points) of $\rho_\pm(t)$.  For $\eta < \eta_{\rm crit}^k$, the minimum of $t(\rho)$ is on the symmetry axis (the top of the caustic), while for $\eta > \eta_{\rm crit}^k$, the minimum occurs off-axis (at $t_0 = t_{\min}^{(k)}$), and $\rho=0$ is a maximum of $t(\rho)$.  The curve given by the irregular points of the parametrization $\bigl[ t_0(\eta), \rho_0(\eta) \bigr]$ (green) has the shape characteristic for a ``pitchfork'' bifurcation.

\subsubsection{Creation of a closed orbit: $t_0 = \frac\eta2 = (k - \frac12)\pi$}
\label{subsubsec:Caustic3a.3}

Here, cases (2b) and (2c) hold simultaneously.  At this irregular point, the return time for the antiparallel closed orbit $\Delta T = \eta = (2k-1)\pi$ becomes an odd multiple of the cyclotron period.  This is exactly the return time for the $k$th snake-type closed orbit which detaches from the ``uphill'' closed orbit at this value of $\eta$ \cite{Peters1994a}.  Because a switch from the ($+$) solution to the ($-$) solution takes place at the endpoint of a snake closed orbit (see the previous discussion of case 2b), the parametrization of the caustic increases in complexity when $\eta$ exceeds the critical value $(2k-1)\pi$.  (See also Fig.~\ref{fig:Caustic3.2}.)  The parabolic shape near the top of the caustic is not affected, however.


\end{document}